\documentclass[useAMS,usenatbib,twocolumn]{mn2e}
\usepackage{epsfig}
\usepackage{amsmath}
\newcommand{\be}{\begin{equation}}
\newcommand{\beq}{\begin{equation}}
\newcommand{\ba}{\begin{eqnarray}}
\newcommand{\ee}{\end{equation}}
\newcommand{\eeq}{\end{equation}}
\newcommand{\ea}{\end{eqnarray}}

\def\lsim{~\rlap{$<$}{\lower 1.0ex\hbox{$\sim$}}}

\def\gsim{~\rlap{$>$}{\lower 1.0ex\hbox{$\sim$}}}

\title[The spatial distribution of neutral hydrogen as traced by low HI mass galaxies]
      {The spatial distribution of neutral hydrogen as traced by low HI mass galaxies}
      \author[Han-Seek Kim et al.]
      {Han-Seek~Kim$^{1}$\thanks{hansikk@unimelb.edu.au}, J. Stuart. B.~Wyithe$^{1,2}$, C. M.~Baugh$^{3}$, C. d. P.~Lagos$^{2,4}$,\vspace{0.3cm} \\ 
      \LARGE\rm{C.~Power$^{2,4}$, and Jaehong~Park$^{1}$} \\
       $^1$School of Physics, The University of Melbourne, Parkville, VIC 3010, Australia\\
       $^2$Australian Research Council Centre of Excellence for All-sky Astrophysics (CAASTRO), 44 Rosehill Street Redfern, NSW 2016, Australia\\
       $^3$Institute for Computational Cosmology, Department of Physics, University of Durham, South Road, Durham DH1 3LE, UK\\
       $^4$International Centre for Radio Astronomy Research, 
                University of Western Australia, 35 Stirling Highway, 
                 Crawley, WA 6009, Australia}
  	
\date{}

\pagerange{\pageref{firstpage}--\pageref{lastpage}}
\pubyear{2016} 

\begin{document}

\maketitle

\label{firstpage}

\begin{abstract}
The formation and evolution of galaxies with low neutral atomic hydrogen (HI) masses, M$_{\rm HI}$$<$10$^{8}h^{-2}$M$_{\odot}$, are affected by host dark matter halo mass and photoionisation feedback from the UV background after the end of reionization. We study how the physical processes governing the formation of galaxies with low HI mass are imprinted on the distribution of neutral hydrogen in the Universe using the hierarchical galaxy formation model, GALFORM. 
We calculate the effect on the correlation function of changing the HI mass detection threshold at redshifts $0 \le z \le 0.5$. We parameterize the clustering as $\xi(r)=(r/r_{0})^{-\gamma}$ and we find that including galaxies with M$_{\rm HI}$$<$10$^{8}h^{-2}$M$_{\odot}$ increases the clustering amplitude $r_{0}$ and slope $\gamma$ compared to samples of higher HI masses. This is due to these galaxies with low HI masses typically being hosted by haloes with masses greater than 10$^{12}{h}^{-1}$M$_{\odot}$, and is in contrast to optically selected surveys for which the inclusion of faint, blue galaxies lowers the clustering amplitude. We show the HI mass function for different host dark matter halo masses and galaxy types (central or satellite) to interpret the values of $r_{0}$ and $\gamma$ of the clustering of HI-selected galaxies.  
We also predict the contribution of low HI mass galaxies to the 21cm intensity mapping signal. We calculate that a dark matter halo mass resolution better than $\sim$10$^{10}{h}^{-1}$M$_{\odot}$ at redshifts higher than 0.5 is required in order to predict converged 21cm brightness temperature fluctuations.
\end{abstract}  

\begin{keywords}
   galaxies: formation -- evolution -- large-scale structure of the Universe -- radio lines: galaxies
   \end{keywords}

\section{Introduction}

In hierarchical structure formation theory, the distribution of galaxies is related to the distribution of dark matter in the Universe. Studies of the spatial distribution of galaxies as a function of colour, luminosity, type and stellar mass therefore give important information about where galaxies form and which halos host them \cite[]{Brown2000, Norberg2002, Madgwick2003, Loh2010, zehavi2011, campbell2015}. 

Neutral hydrogen is a key ingredient of galaxy formation as it traces the processes of gas accretion, outflow from galaxies and star formation. It is therefore important to understand the relation between neutral hydrogen and dark matter haloes. However, understanding galaxy formation and evolution using neutral hydrogen observations is hampered by the weakness of the signal compared to optical observations. The last decade has seen improvements in both the volume and flux limit of surveys that probe the HI content of galaxies from the HI Parkes All Sky Survey \cite[HIPASS,][]{zwaan.etal.2005}, the Arecibo Legacy Fast ALFA \cite[ALFALFA,][]{martin.etal.2010} and the Arecibo Ultra-Deep Survey \cite[AUDS,][]{hoppmann2015}. Studies of the distribution of HI-selected galaxies in the local Universe have shown that these galaxies are an unbiased (weakly clustered) galaxy population \cite[]{Meyer2007, Martin2012, papastergis2013}.

There have been previous studies of the large scale structure of galaxies based on their neutral hydrogen using the halo occupation distribution (HOD) formalism \cite[]{wyithe2009} and hierarchical galaxy formation models \cite[]{Kim2011, Kim2013a}. These works have focused on high HI mass galaxies (M$_{\rm HI}$$>$ 10$^{9.25}h^{-2}$M$_{\odot}$), which are mainly central galaxies located in dark matter haloes of masses $\sim$ 10$^{12}h^{-1}$M$_{\odot}$. In contrast, red massive galaxies have either consumed most of their gas (neutral hydrogen) in forming stars or have expelled it by feedback processes \cite[]{Power2010,Kim2011,Lagos2011}, as shown by their very low neutral gas content \cite[]{Young2011,Serra2012,Boselli2014b}. HI deficient galaxies also result from ram pressure stripping by hot gaseous atmospheres around satellite galaxies in groups and clusters, and tidal interactions between galaxies in high density environments \cite[]{MCC2008,FONT2008,Cortese2012,Boselli2014a,bahe2015}. These processes all affect the distribution of satellite galaxies in groups and clusters. 

\cite{Kim2015HI} demonstrated the importance of photoionisation feedback for galaxy formation and evolution in low mass dark matter haloes, and the low-mass end of the HI mass function in the local Universe \cite[see also][]{Lagos2011,Kim2013a}. \cite{Kim2015HI} showed that the contribution of low mass dark matter haloes to the low-mass end of the HI mass function changes dramatically when different photoionisation feedback modelling schemes are applied. Therefore, it is natural to expect these models to affect the predicted clustering of low HI mass galaxies.

The number of individual galaxies detected in HI will greatly increase through ongoing and upcoming large volume HI-selected galaxy surveys using the SKA \cite[Square Kilomotre Array; e.g.][]{Baugh2004SKA,Power2010,Kim2011} and its pathfinders, such as ASKAP \cite[Australian Square Kilometre Array Pathfinder; cf.][]{ASKAP2008}, and MeerKAT \cite[Meer Karoo Array Telescope; cf.][]{MeerKAT2007} in the southern hemisphere, FAST (Fivehundred-meter Aperture Spherical Telescope)\footnote{details at http://fast.bao.ac.cn/en/}, and APERTIF (APERture Tile In Focus)\footnote{details at www.astron.nl/jozsa/wnshs} in the northern hemisphere. In addition, {the BINGO \cite[the BAO in Neutral Gas Observations;][]{Battye2012}, CHIME \cite[the Canadian Hydrogen Intensity Mapping Experiment;][]{Bandura2014}, and SKA are expected to} detect the distribution of neutral hydrogen at 0.5$<$$z$$<$3 in the Universe using the 21cm intensity mapping technique which does not require individual HI-selected galaxies to be resolved \cite[]{WL07,Wolz2014,Santos2015}. { There have been several theoretical models predicting the HI intensity during the post-reionization era. Each model includes a prescription for the relation between HI and dark matter halo masses. This results in the implicit assumption that the HI traces the underlying matter distribution with a linear bias \citep{Bagla2010,Guha2012,Padman2015,Villa2016}. For example, \cite{Bagla2010} propose phenomenological functions for this relation based on the observation that haloes much more massive than a few times 10$^{11}$M$_{\odot}$ do not host significant amounts of HI  \citep{Catinella2013, Denes2014}. This simple model successfully explains both observations of the abundance of damped lyman alpha absorbers at redshifts between 2.4 to 4, and the bias of HI-selected galaxies in the local Universe \citep{Villa2014, Padman2015, Sarkar2016}. On the other hand, using zoom hydrodynamic simulations, \cite{Villa2016} found the mass in neutral hydrogen of dark matter halos increases monotonically with halo mass, and can be well described by a power-law.} Here, we show the importance of the contribution of galaxies with low HI masses to the large scale distribution of neutral hydrogen in the Universe{ , and why they cannot be neglected if an accurate measurement the HI intensity mapping power spectrum is desired.}
 
The structure of the paper is as follows. In \S\ref{sec:model}, we provide
a brief overview of the hierarchical galaxy formation models used.
In \S\ref{sec:HICLUSTERING} we show the clustering of HI selected galaxies from the models and the comparison between the models and observations. In \S\ref{sec:HIGHCLUSTERING} we extend predictions of HI galaxy clustering to higher redshifts and show predictions for 21cm brightness fluctuations at different redshifts in \S\ref{sec:INTENSITY}. We finish with a summary of our results in \S\ref{sec:summary}. 

\section{The Galaxy Formation Model} 
\label{sec:model}

We use the semi-analytical galaxy formation model {GALFORM} 
\citep[cf.][]{cole.etal.2000,baugh.2006,Lacey2015} described in \citet{Lagos2012} and \cite{Kim2015HI} 
to predict the properties of galaxies which form and evolve in the
$\Lambda$CDM cosmology. 
GALFORM models the key physical 
processes involved in galaxy formation, including the gravitationally driven assembly 
of dark matter halos, radiative cooling of gas and its collapse to form 
centrifugally supported discs, star formation, feedback from
supernovae (SNe) and active galactic nuclei (AGN), and photoionisation. We implement GALFORM 
in halo merger trees extracted from the Millennium-II cosmological N-body simulation\footnote{The cosmological
parameters adopted for the Millennium Simulations are a total matter density
$\Omega_{\rm M}=0.25$, baryon density $\Omega_{\rm b}=0.045$, 
 dark energy density $\Omega_{\Lambda}=0.75$, Hubble parameter
$H_{0}=73 \,{\rm kms^{-1}}\,{\rm Mpc}^{-1}$, the primordial scalar spectral 
index $n_{\rm s}=1$ and the fluctuation amplitude $\sigma_{8}=0.9$.} which has a 100$h^{-1}$Mpc box size \cite[][]{MII2009}. 

{ We use the Dhalo merger trees \cite[D-Trees;][]{Jiang2014}. These were designed for galaxy formation modelling and form the basis of the Durham semi-analytic galaxy formation model, GALFORM. The D-Trees algorithm is based on the Friends-Of-Friends algorithm, FOF, \citep{Davis1985}. The FOF haloes are decomposed into subhaloes by SUBFIND \citep{Springel2005G}. In the D-Trees, subhaloes 
have consistent membership over time in the sense that once a subhalo is accreted by a Dhalo it never demerges. In contrast to the FOF haloes, in this process we also split some FOF haloes into two or more Dhaloes when SUBFIND substructures are well separated and only linked into a single FOF halo by bridges of low density material. The central is associated with the main subhalo. Satellite galaxies were originally centrals until their FOF halo merged with a larger one, and are associated with the corresponding subhalo, or the most bound particle from this subhalo if the subhalo itself can no longer be identified. Full details of the Dhalos can be found in \cite{Merson2012} and \cite{Jiang2014}.} 

The particle mass of the simulation is 6.89$\times$10$^{6}h^{-1}$M$_{\odot}$ and we retain haloes down to 20 particles. { In the extreme case where all available baryons are in HI, the HI mass of a galaxy in such a halo would be $\sim$ 1.8$\times$10$^{6}$$h^{-2}$M$_{\odot}$. Thus the resolution limit of the Millennium-II simulation is sufficient to model the HI mass function over the range of current observations (i.e., down to M$_{\rm HI}$$\sim$10$^{6}$M$_{\odot}$).}

\citet{Lagos2011} extended GALFORM by modelling the splitting
of cold gas in the interstellar medium (ISM) into its HI and H$_2$ components 
and by explicitly linking star formation to the amount of H$_2$ present in a 
galaxy. \citet{Lagos2011} compared empirically and theoretically 
derived star formation laws \citep[cf.][]{blitz.2006,krumholz.etal.2009} with 
a variety of observations and found that the
empirical law of \citet{blitz.2006} (see also \citealt{leroy.etal.2008}) is 
favoured by these data. 

Another important physical process affecting HI masses in galaxies is reionization. The standard reionizaton model in GALFORM suppresses cooling in galaxies with host halo circular velocity $V_{\rm circ}$ below a threshold $V_{\rm cut}$ below redshift $z_{\rm cut}$ \cite[see][]{benson.etal.2002}. \cite{Lagos2011b} adopted $V_{\rm cut}$=30 km/s and $z_{\rm cut}$=10. \cite{Lagos2012} changed two parameters corresponding to the star formation timescale during starbursts, $\tau_{\rm min}$=100 Myr and $f_{\rm dyn}$=50 when the star formation timescale is calculated as $\tau_{SF}$=min($f_{\rm dyn}$ $\times$ $\tau_{\rm dyn, bulge}$, $\tau_{\rm min}$), to better match the observed rest-frame UV luminosity function over the redshift range z=3-6 (hereafter Lagos2012). Here $\tau_{\rm dyn, bulge}$ is the dynamical timescale of the bulge.

\citet{Kim2015HI} introduced a new model for photoionisation feedback in order to correctly describe the low-mass end of the HI mass function (10$^{6}h^{-2}\rm M_{\odot}$$<$$M_{\rm HI}$$<$$10^{8}h^{-2}\rm M_{\odot}$) estimated from the ALFALFA survey at $z$=0 (hereafter Kim2015). 
In contrast to Lagos2012, \cite{Kim2015HI} adopted the redshift-dependent photoionisation feedback model \citep{dijkstra2004} described in \cite{sobacchi.etal.2013}. \cite{Kim2015HI} use a redshift dependent $V_{\rm cut}$ given by:  
\begin{equation}
\begin{split}
V_{\rm cut}(z)  [{\rm km/s}] = V_{\rm cut0}(1+z)^{\alpha_{v}}
                   \left[1-\left({1+z} \over {1+z_{\rm IN}}\right)^{2}\right]^{2.5/3},
\end{split}
\label{VZ}
\end{equation}
where $z_{\rm IN}$ is the redshift of UV background exposure, $V_{\rm cut0}$ is the circular velocity of dark matter halos at $z$=0 below which photoionisation feedback suppresses gas cooling and $\alpha_{v}$ parameterises the redshift dependence of the velocity cut. 
We use the best fitting set of parameters in \cite{Kim2015HI}, ($V_{\rm cut0}$, $\alpha_{v}$)=(50 km/s,-0.82) at $z_{\rm IN}$=10, which was found by comparing the predicted HI mass function to the measurements by the ALFALFA survey. The predicted HI mass function using this set of parameters shows an improved match with observations compared to the Lagos2012 model. Note that \cite{sobacchi.etal.2013} suggested $\alpha_{v}$=-0.2 and a critical halo mass that is ten times larger that the one we used for a UV background intensity of 1$\times$10$^{21}$erg/s/Hz/cm$^{2}$/sr \cite[for more details see][]{Kim2015HI}.

\section{Large scale distribution of neutral hydrogen in the local Universe}
\label{sec:HICLUSTERING}
We show clustering predictions for galaxies selected using different HI mass thresholds in \S~\ref{sec:HIDep}.
To compare the spatial distributions of galaxies selected by different HI mass criteria, we investigate the contributions of different host dark matter halo masses and types of galaxies to the HI mass function in \S~\ref{sec:Under}.

We quantify the large scale distribution of galaxies using the two point correlation function. This is written in terms of the excess probability, compared to a random distribution, of finding another galaxy at a separation $r$, 
\begin{equation}
dP=\bar{n}^{2}[1+\xi(r)]\delta V_{1} \delta V_{2},
\end{equation}
where $\bar{n}$ is the mean number density of the galaxy sample and $\delta V_{1}$, $\delta V_{2}$ are volume
elements. In the case of an HI-selected galaxy catalogue produced from a galaxy formation model combined
with a periodic N-body simulation, we can measure the correlation function using
\begin{equation}
1+\xi(r)=D_{\rm HI}D_{\rm HI}/(\bar{n}_{\rm HI}^{2}V dV),
\end{equation}
where $D_{\rm HI}D_{\rm HI}$ is the number of pairs of HI-selected galaxies with separations in the range $r$ to $r$+$\delta r$,  $\bar{n}_{\rm HI}$ is mean number density of HI-selected galaxies, $V$ is the volume of the simulation, and $dV$ is the differential volume corresponding to $r$ to $r$+$\delta r$. 

The correlation function can be described by a  power law using two parameters as
\begin{equation}\label{powerlaw}
\xi(r)=(r/r_{0})^{-\gamma},
\end{equation}
where $r_{0}$ indicates the correlation length at which $\xi(r_{0})$=1, corresponding to twice the probability relative to a random distribution. The value of $r_{0}$ is related to the typical host dark matter halo mass of the selected galaxies, and $\gamma$ is related to the number of satellite galaxies in a host dark matter halo. { Here, the parameters $r_{0}$ and $\gamma$ are derived by { minimizing $\chi^{2}$} over separations between 1 and 10$h^{-1}$Mpc, and considering $\gamma$ values between 0 and 5.}

\subsection{HI mass dependent clustering}\label{sec:HIDep} 

The ALFALFA survey measures the HI mass function down to HI masses of $\sim$10$^{6}h^{-2}$M$_{\odot}$. This low mass limit provides an important test for galaxy formation models. The Kim2015 model is able to reproduce the observed HI mass function from ALFALFA by adopting a redshift-dependent photoionisation feedback model (see \S\ref{sec:model} and also Fig.~\ref{HankPHZ}). 

\begin{figure}
  \includegraphics[width=8.cm]{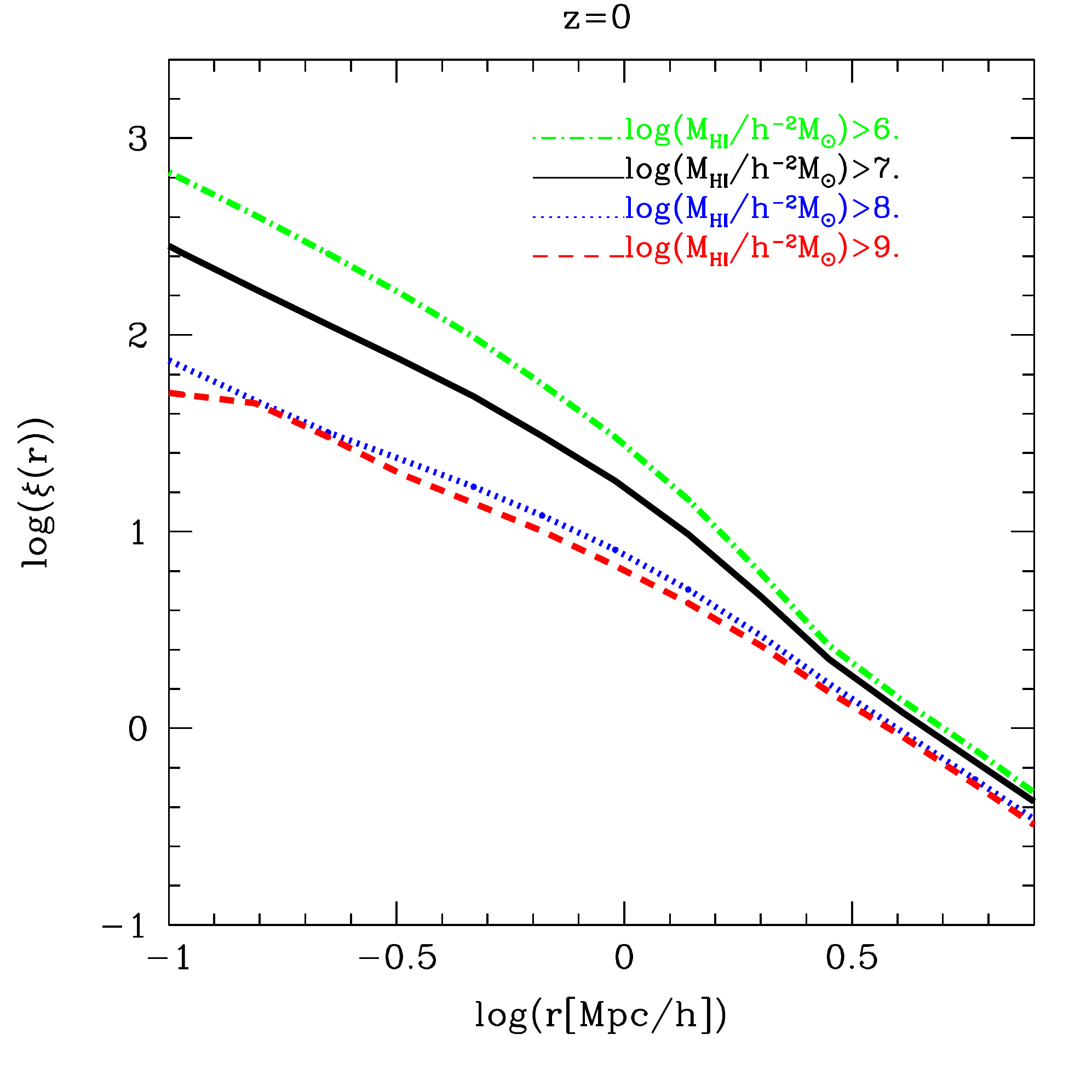}
  \caption{The real space correlation function at $z$=0 predicted for different HI mass thresholds using the Kim2015 model, as labelled. { Reprinted from \citet{Kim2015HI}}.}
  \label{CFModels}
\end{figure}

In Fig.~\ref{CFModels}, we show clustering predictions for galaxies selected using different HI mass thresholds. The clustering predictions for the two highest HI mass thresholds (M$_{\rm HI}$$>$10$^{8}h^{-2}$M$_{\odot}$ and M$_{\rm HI}$$>$10$^{9}h^{-2}$M$_{\odot}$) are very similar, both in shape ($\gamma$) and the correlation length ($r_{0}$). However, the two lowest HI mass thresholds (M$_{\rm HI}$$>$10$^{6}h^{-2}$M$_{\odot}$ and M$_{\rm HI}$$>$10$^{7}h^{-2}$M$_{\odot}$) show larger correlation lengths (larger $r_{0}$) and also steeper correlation functions (larger $\gamma$) than the two high HI mass samples (see values of $r_{0}$ and $\gamma$ in Table~\ref{Tabler0} and Fig.~\ref{HIBP} at $z$=0). Overall, the trend of clustering as a function of HI mass threshold shows that galaxies become more clustered as the HI mass threshold {\it decreases}.    

\subsection{Understanding the HI mass-dependent clustering in the Kim2015 model }\label{sec:Under}

The clustering of galaxy samples can be broken down into the contribution of the host dark matter haloes \cite[]{berlind2002,Kravtsov2004,Kim2011,Kim2012}. We therefore investigate the contributions to the HI mass function from different galaxies and different mass host dark matter haloes.

\begin{figure*}
\vspace{0cm}
  \includegraphics[width=8.cm]{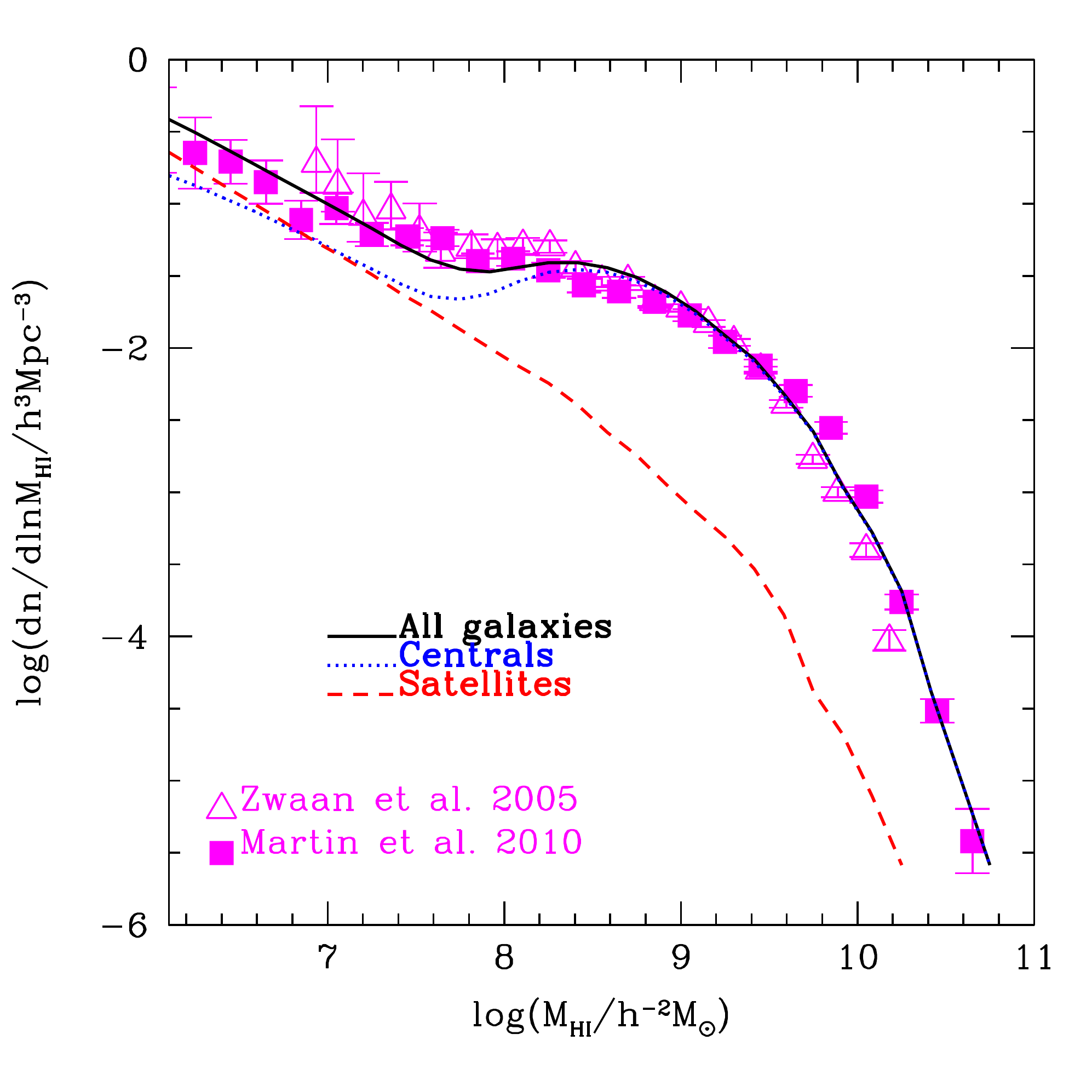}
   \includegraphics[width=8.cm]{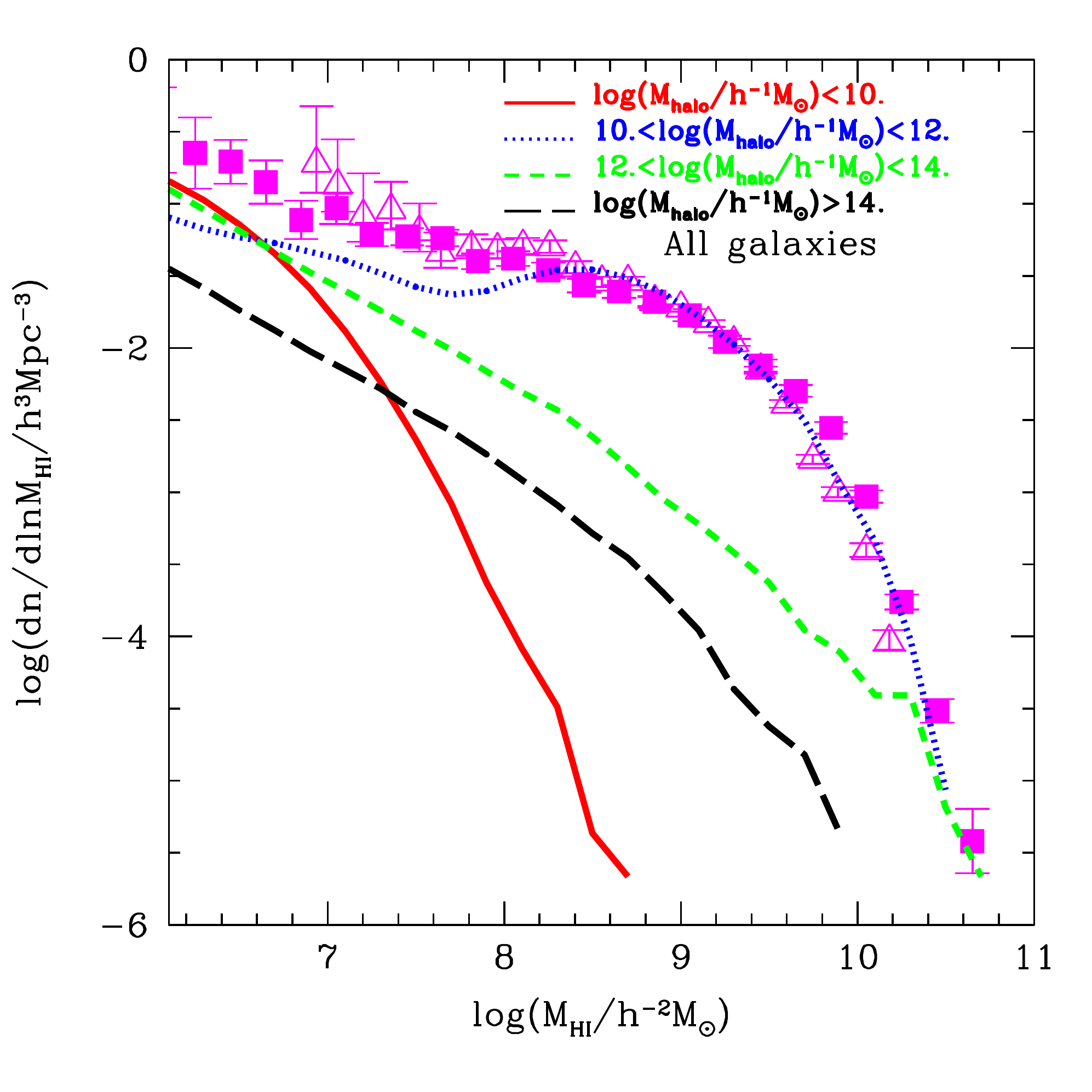}
   \includegraphics[width=8.cm]{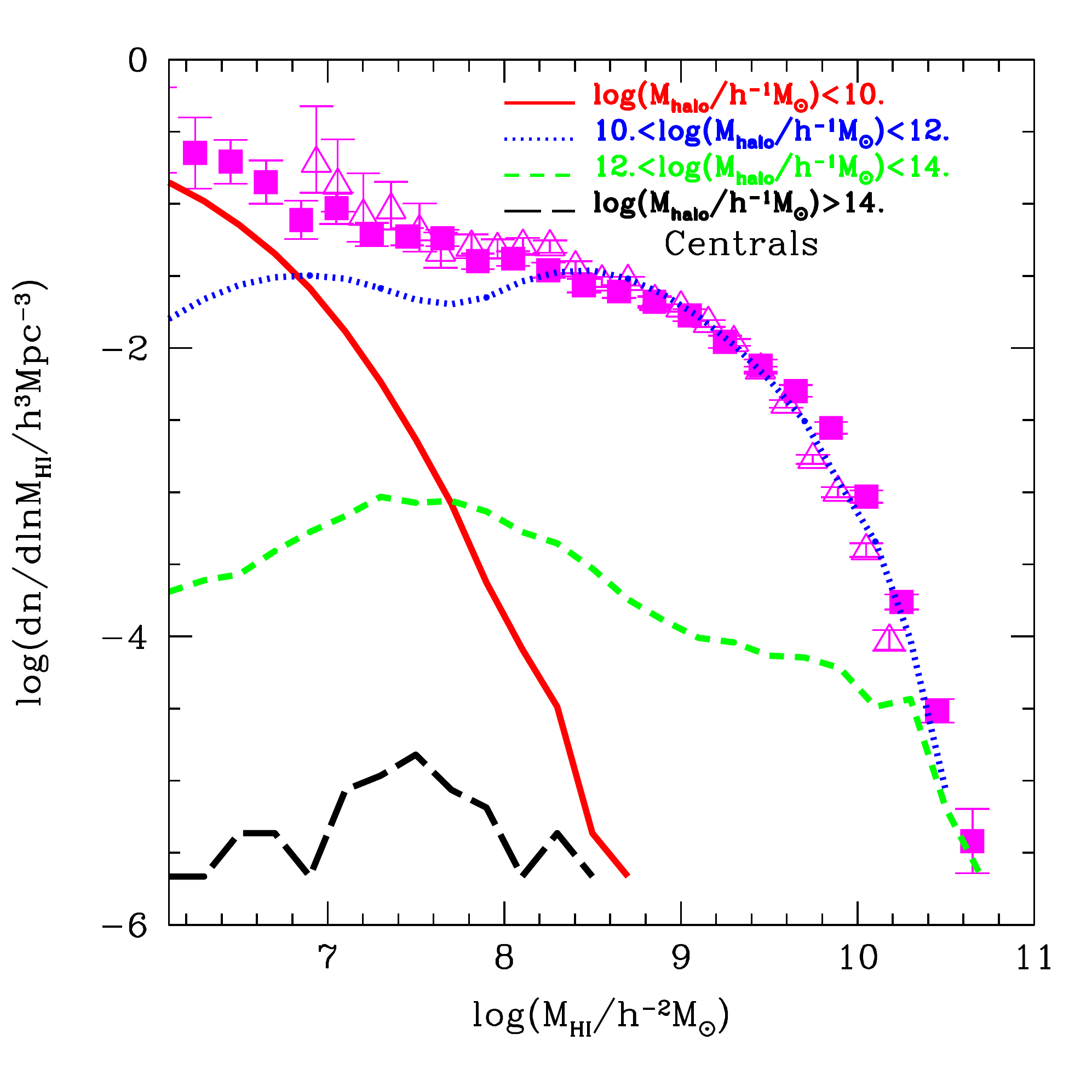}
\includegraphics[width=8.cm]{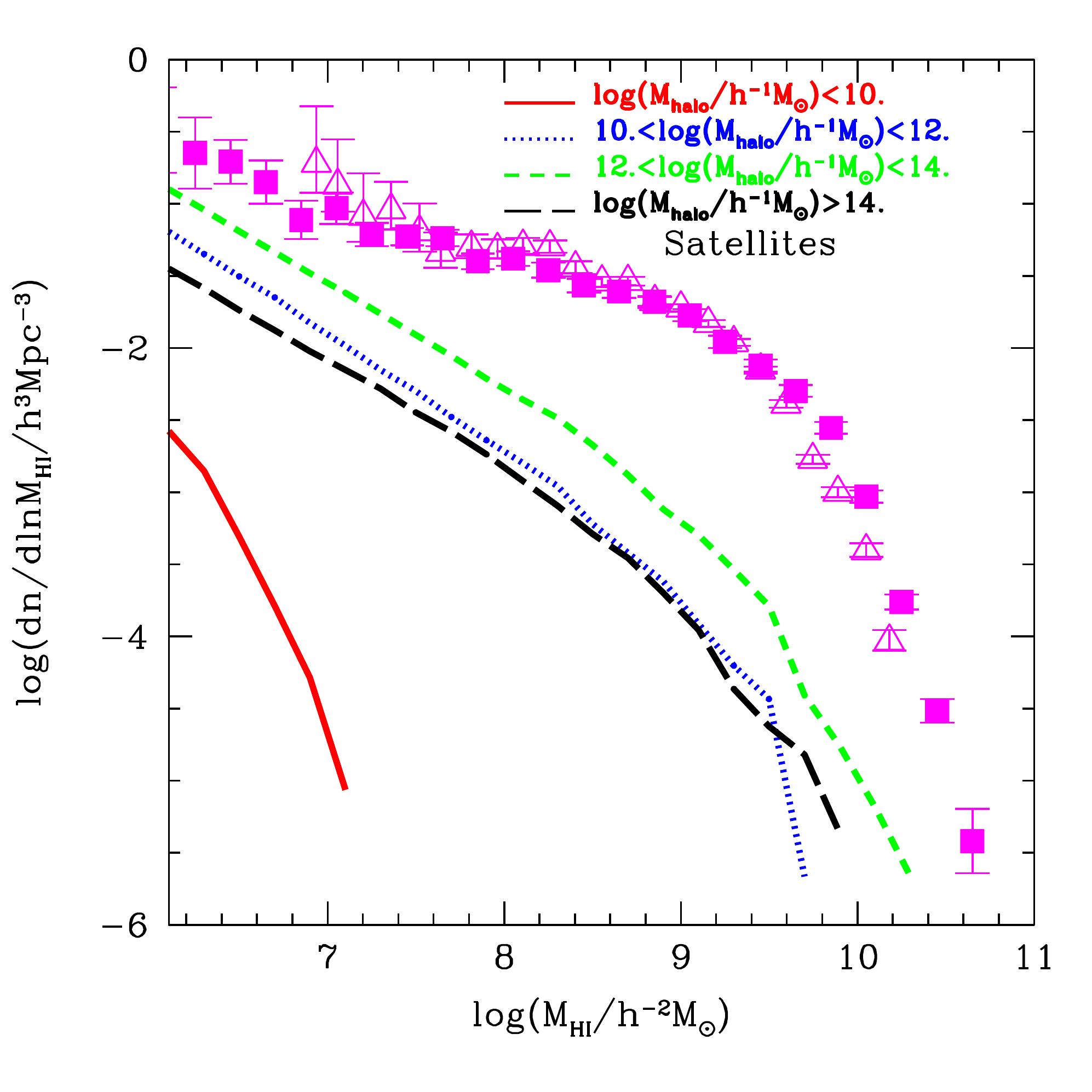}
  \caption{The predicted HI mass function in the Kim2015 model. The top-left panel shows the contribution of central and satellite galaxies to the HIMF ({ Originally plotted in \citet{Kim2015HI}}). The top-right panel shows the contribution of different masses of { host} dark matter haloes. The bottom panels show the contribution of different masses of { host} dark matter haloes for central galaxies (left panel) and satellite galaxies (right panel).
    The symbols correspond to observational estimations from the local Universe from 
    HIPASS \citep[open triangles; cf.][]{zwaan.etal.2005} and ALFALFA
    \citep[filled squares; cf.][]{martin.etal.2010}. { M$_{\rm halo}$ is the mass of the associated Dhalo.}}
  \label{HankPHZ}
\end{figure*}

The top-left panel of Fig.~\ref{HankPHZ} shows the contribution of different types of HI-selected galaxies (black solid line for all galaxies, blue dotted line for central galaxies, and red dashed line for satellite galaxies) to the HI mass function. Central galaxies are the dominant (over 90\%) type of galaxy when we focus on galaxies with HI masses $M_{\rm HI}$$>$10$^{8.5}h^{-2}$M$_{\odot}$. However, the contributions of central galaxies and satellite galaxies to the HI mass function are similar for $M_{\rm HI}$$<$10$^{8}h^{-2}$M$_{\odot}$. { This is consistent with the findings from hydrodynamic simulations \citep{Dave2013}.}

The top-right panel of Fig.~\ref{HankPHZ} shows the HI mass functions for galaxies in different bins of host dark matter halo mass. { Note that the host dark matter halo mass, M$_{\rm halo}$, is the mass of the associated host Dhalo.}
Most galaxies with HI masses $>$10$^{7.5}h^{-2}$M$_{\odot}$ reside in { host} dark matter haloes with masses between 10$^{10}h^{-1}$M$_{\odot}$ and 10$^{12}h^{-1}$M$_{\odot}$. 
{ Host} dark matter haloes with mass $<$10$^{10}h^{-1}$M$_{\odot}$ mainly contribute to the low-mass end of the HI mass function, $M_{\rm HI}$$<$10$^{7}h^{-2}$M$_{\odot}$, while making a negligible contribution to the HI mass function at HI masses higher than 10$^{8}h^{-2}$M$_{\odot}$. The most massive { host} dark matter haloes ($>$10$^{14}h^{-1}$M$_{\odot}$) contribute less than 10$\%$ to the abundance of galaxies at all HI masses. 

We show { host} dark matter halo contributions divided into central (bottom-left panel of Fig.~\ref{HankPHZ}) and satellite galaxies (bottom-right panel of Fig.~\ref{HankPHZ}). For central galaxies, the dominant contribution to the HI mass function at $M_{\rm HI}$$>$10$^{7.5}h^{-2}$M$_{\odot}$ is from { host} dark matter haloes with masses 10$^{10}h^{-1}$M$_{\odot}$$<$$M_{\rm halo}$$<$10$^{12}h^{-1}$M$_{\odot}$. Central galaxies which reside in higher mass host dark matter haloes (10$^{12}h^{-1}$M$_{\odot}$$<$$M_{\rm halo}$$<$10$^{14}h^{-1}$M$_{\odot}$ and $>$10$^{14}h^{-1}$M$_{\odot}$) represent a very small fraction of HI-selected galaxies. { Host d}ark matter haloes with masses less than 10$^{10}h^{-1}$M$_{\odot}$ mainly host central galaxies with $M_{\rm HI}$$<$10$^{7}h^{-2}$M$_{\odot}$. 

{ Photoionization feedback in
the model suppresses the cooling of gas in central galaxies which have circular velocities less than V$_{\rm cut}$. Therefore the suppressed cooling of gas in central galaxies introduces the departure from a Schechter-like function. 
On the other hand,
the abundance of satellite galaxies in massive dark matter haloes increases monotonically
towards low HI mass. The reason why the HIMF of satellite galaxies does not show a dip at low HI masses is that the satellites were mainly formed
before reionization.}

We find that the number of HI-selected galaxies that are satellites increases with decreasing HI mass for all host dark matter halo masses. Furthermore, at $M_{\rm HI}$$<$10$^{7}h^{-2}$M$_{\odot}$, HI satellite galaxies dominate the number density. These satellite galaxies tend to lie in { host} dark matter haloes of mass M$_{\rm halo}$$>$10$^{12}h^{-1}$M$_{\odot}$.

In Fig.~\ref{CFModels}, we found that the correlation length $r_{0}$ becomes larger as the HI mass threshold decreases for the selected galaxy samples. This is due to the larger contributions of high mass host dark matter haloes (M$_{\rm halo}$$>$10$^{12}h^{-1}$M$_{\odot}$) to the HI mass function (see top-right panel of Fig.~\ref{HankPHZ}) at  M$_{\rm HI}$$<$10$^{8}h^{-2}$M$_{\odot}$.  
In addition, the contribution of satellite galaxies residing in the highest mass host dark matter haloes (M$_{\rm halo}$$>$10$^{14}h^{-1}$M$_{\odot}$) at M$_{\rm HI}$$<$10$^{8}h^{-2}$M$_{\odot}$ leads to a boost in the clustering amplitude at small separations (see Fig.~\ref{CFModels}). We also find that the slope of the correlation function decreases with increasing HI mass due to the reduced contribution of satellite galaxies.

\begin{figure*}
    \includegraphics[width=8.cm]{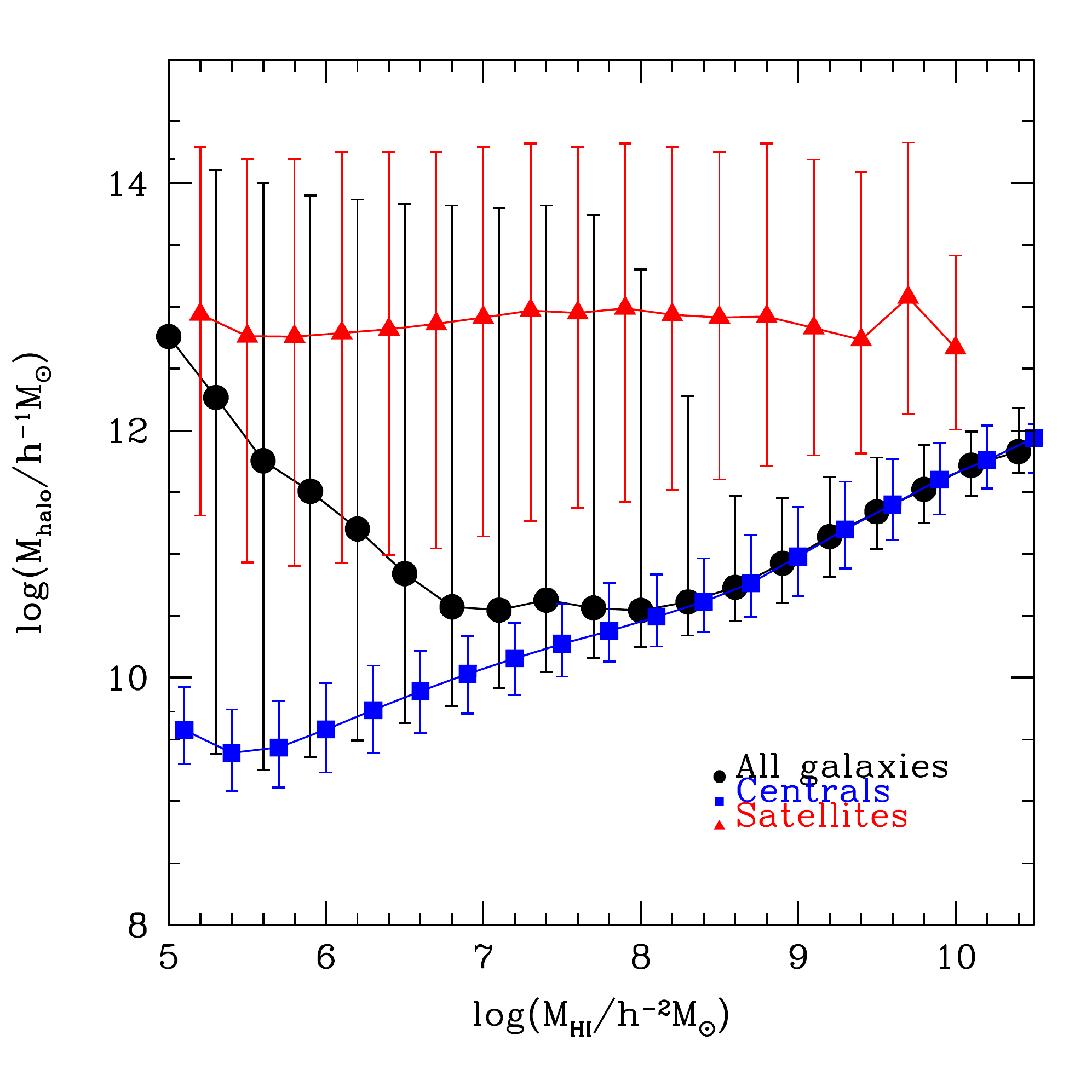}\vspace{-1cm}
    \includegraphics[width=8.cm]{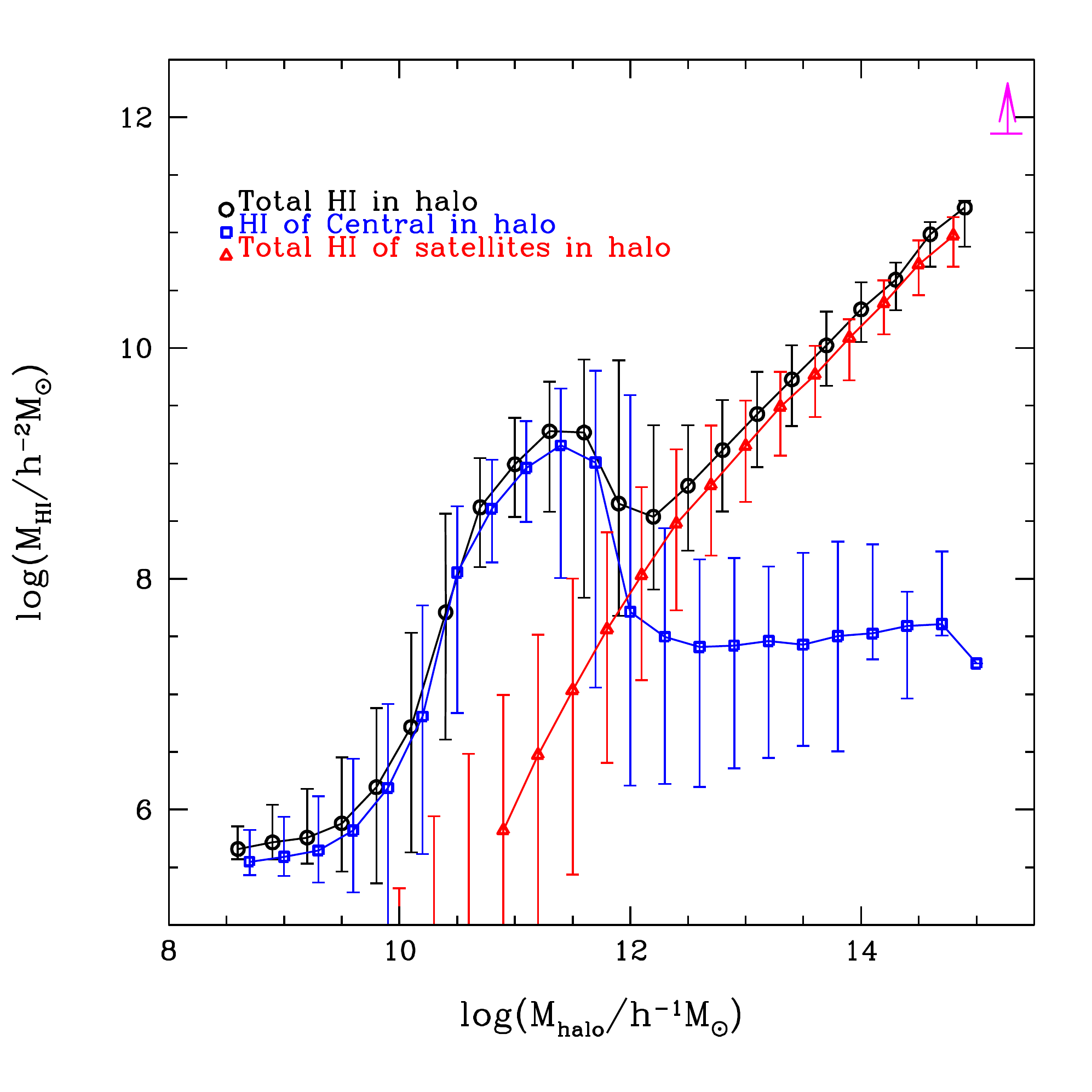}
    \includegraphics[width=8.cm]{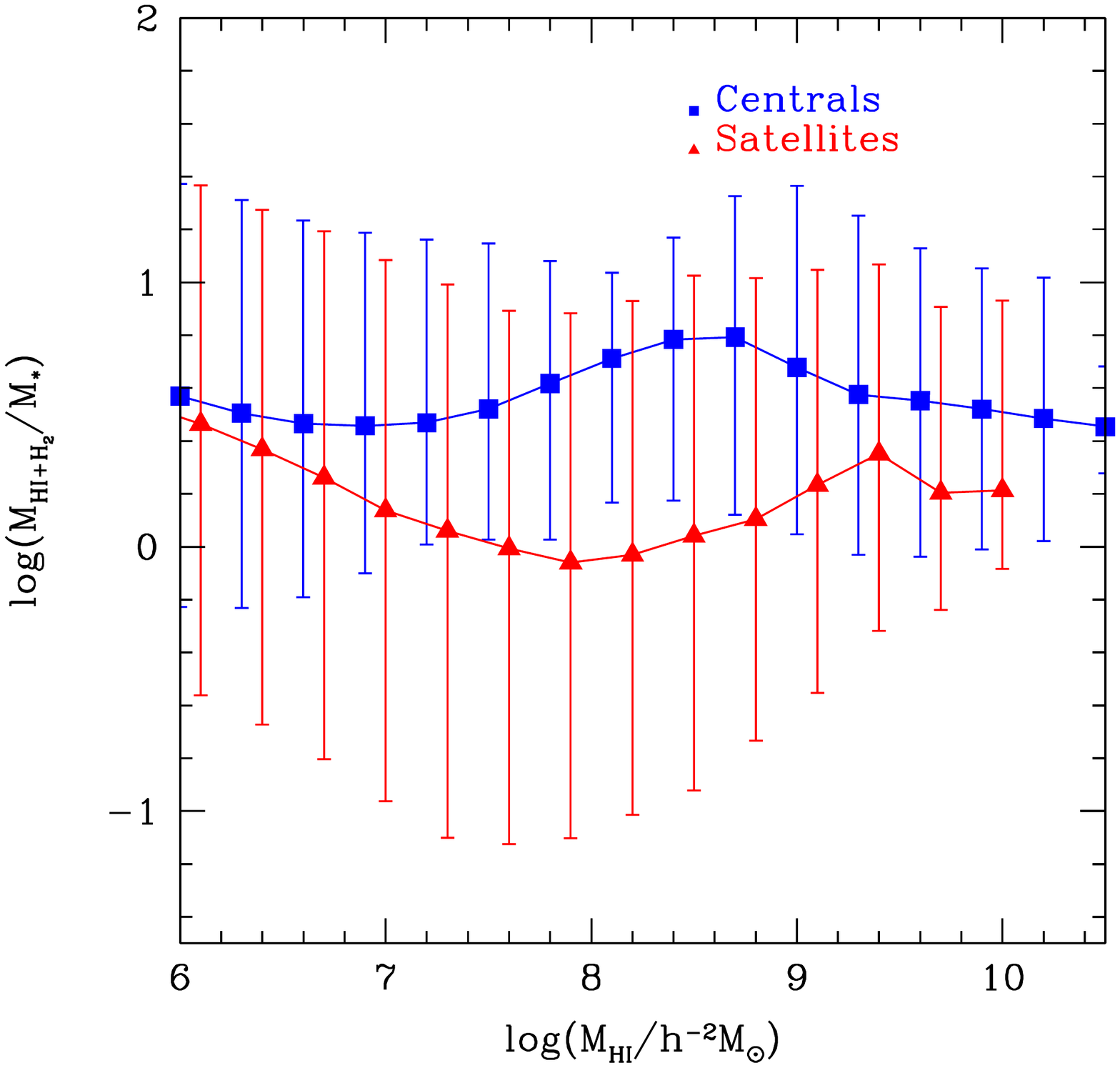}
    \vspace{-2cm}
  \caption{The {top} left panel shows the host halo mass plotted in bins of HI mass {at z=0}. The symbols show the median dark matter mass as
a function of the HI mass of galaxies, for central galaxies (blue squares), satellite galaxies (red triangles)
and all galaxies (black circles). The bars show the 10th-90th percentile ranges of the
distributions of {host} dark matter halo masses. The {top} right panel shows the {\it total} HI mass contained
in a halo {at z=0}. The symbols show the median total HI mass as
a function of {host} dark matter halo mass, for central galaxies (open blue squares), all satellite galaxies (open red triangles)
and all galaxies (open black circles) in a halo. The bars show the 10th-90th percentile ranges of the
distributions of total HI masses. In this figure we include all galaxies which have HI mass larger than 10$^{4}h^{-1}$M$_{\odot}$. { The magenta arrow represents the lower limit of the overall HI mass of the Coma cluster by summing HI masses of 223 HI detected late type galaxies in \citet{Gavazzi2006}. The total mass of the Coma cluster from \citet{Kubo2007} is shown for comparison.} { M$_{\rm halo}$ is the mass of the associated Dhalo.} {The bottom panel shows the neutral hydrogen gas fraction in bins of HI mass. The symbols show the median neutral hydrogen gas ($M_{\rm HI+H_{2}}$) fraction as
a function of the HI mass of galaxies, for central galaxies (blue squares), and satellite galaxies (red triangles). $M_{\rm H_{2}}$ is the molecular hydrogen mass and $M_{\star}$ is the stellar mass. The bars show the 10th-90th percentile ranges of the distributions of neutral hydrogen gas fractions.}}
 \label{HIDISMH}
\end{figure*}

We show the distribution of host dark matter halo mass as a function of HI mass in galaxies in the {top} left panel of Fig.~\ref{HIDISMH}. The symbols show the median { host} dark matter halo mass as a function of HI mass for central galaxies (blue squares), satellite galaxies (red triangles)
and all galaxies (black circles). The bars show the 10th-90th percentile range of the distribution of host dark matter halo masses. The host dark matter halo mass of central galaxies increases as the HI mass increases. In contrast, the median host dark matter halo mass for satellite HI-selected galaxies is constant over the entire range of HI masses (median value $\sim$10$^{13}h^{-1}$M$_{\odot}$). 

We also show in the {top} right panel of Fig.~\ref{HIDISMH} the {\it total} HI mass contained in all galaxies within
a halo. The symbols show the median total HI mass as
a function of { host} dark matter halo mass, for central galaxies (open blue squares), all satellite galaxies (open red triangles)
and all galaxies (open black circles) in a halo. The bars show the 10th-90th percentile range of the
distribution of total HI masses. The total HI mass in a halo is contributed mainly by the central galaxy for { host} dark matter haloes less massive than 10$^{12}h^{-1}$M$_{\odot}$ and satellite galaxies for { host} dark matter haloes more massive than 10$^{12}h^{-1}$M$_{\odot}$. Note that these measurements have not been made observations yet and thus these are model predictions that need to be verified by future observations. In this model there is very little HI gas in the central galaxy haloes more massive than this due to the shut down of the cooling flow by AGN heating \cite[][]{Kim2011}. The total HI mass of satellites (adding all HI masses of satellites together in a halo) increases as the halo mass increases. This predicted relation between total HI mass in a { host} dark matter halo and dark matter halo mass would be the physically motivated input to make HI intensity mapping mock observations based on dark haloes using large volume dark matter only simulations. {The bottom panel of Fig.~\ref{HIDISMH} shows the neutral hydrogen gas fraction, $M_{\rm HI+H_2}/M_{\star}$, in bins of HI mass. $M_{\rm H_{2}}$ is the molecular hydrogen mass and $M_{\star}$ is the stellar mass. The symbols show the median neutral hydrogen gas fraction as
a function of the HI mass of galaxies, for central galaxies (blue squares), and satellite galaxies (red triangles). The bars show the 10th-90th percentile ranges of the distributions of neutral hydrogen gas fractions. The overall neutral hydrogen gas fraction for satellites is lower than centrals. Because the hot gas in the subhaloes around satellites is removed instantaneously and is transferred to the hot gas reservoir of the main halo once a galaxy becomes a satellite in the model. This removal of the hot gas prevents the satellite galaxy to accrete newly cooled gas, condemning it to have ever lower gas fractions as star formation continues (see discussion in \citet{Lagos2011} for the competing environment and star formation processes). The feedback effects are also imprinted on the low HI mass end from the suppression of cooling of gas by the photoionization and the high HI mass end from the shutting down of the cooling flow by the AGN feedback for centrals.}

\subsection{Imprinted photoionisation feedback on HI clustering}\label{sec:CoObs}
In the previous subsection we presented predictions for HI galaxy clustering and the mass function. By comparing these predictions with the observed HI mass function and clustering of HI-selected galaxies from HIPASS \cite[]{Meyer2007} and ALFALFA \cite[]{Martin2012} \cite[see][]{Kim2011,Kim2012a,Kim2015HI} we can test the galaxy formation model. \cite{Martin2012} showed that HI-selected galaxies display an anisotropic clustering pattern, and are less clustered than dark matter on scales $<$ 5Mpc. Furthermore, \cite{papastergis2013} found no evidence for HI mass-dependent clustering over the HI mass range between 10$^{8.5}$M$_{\odot}$ and 10$^{10.5}$M$_{\odot}$. \cite{Meyer2007}
used 4315 sources from the HIPASS survey to measure the clustering of HI-selected galaxies. \cite{Martin2012} used a sample of $\sim$10150 galaxies (M$_{\rm HI}$$>$10$^{6.2}h^{-2}$M$_{\odot}$) and \cite{papastergis2013} used a sample of $\sim$6000 galaxies detected by the ALFALFA 21cm survey (M$_{\rm HI}$$>$10$^{7.5}h^{-2}$M$_{\odot}$), to measure the clustering properties of HI-selected
galaxies. Whilst these surveys represent a tremendous step forwards in terms
of blind surveys of the HI content of galaxies, it is important to bear in
mind when considering clustering measurements made from them that
they contain orders of magnitude fewer galaxies and sample smaller volumes
than local optical surveys, such as the two-degree-Field Galaxy Redshift Survey
\citep{2df}, or Sloan Digital Sky Survey \citep{sdss}.

\begin{figure*}
  \includegraphics[width=8.cm]{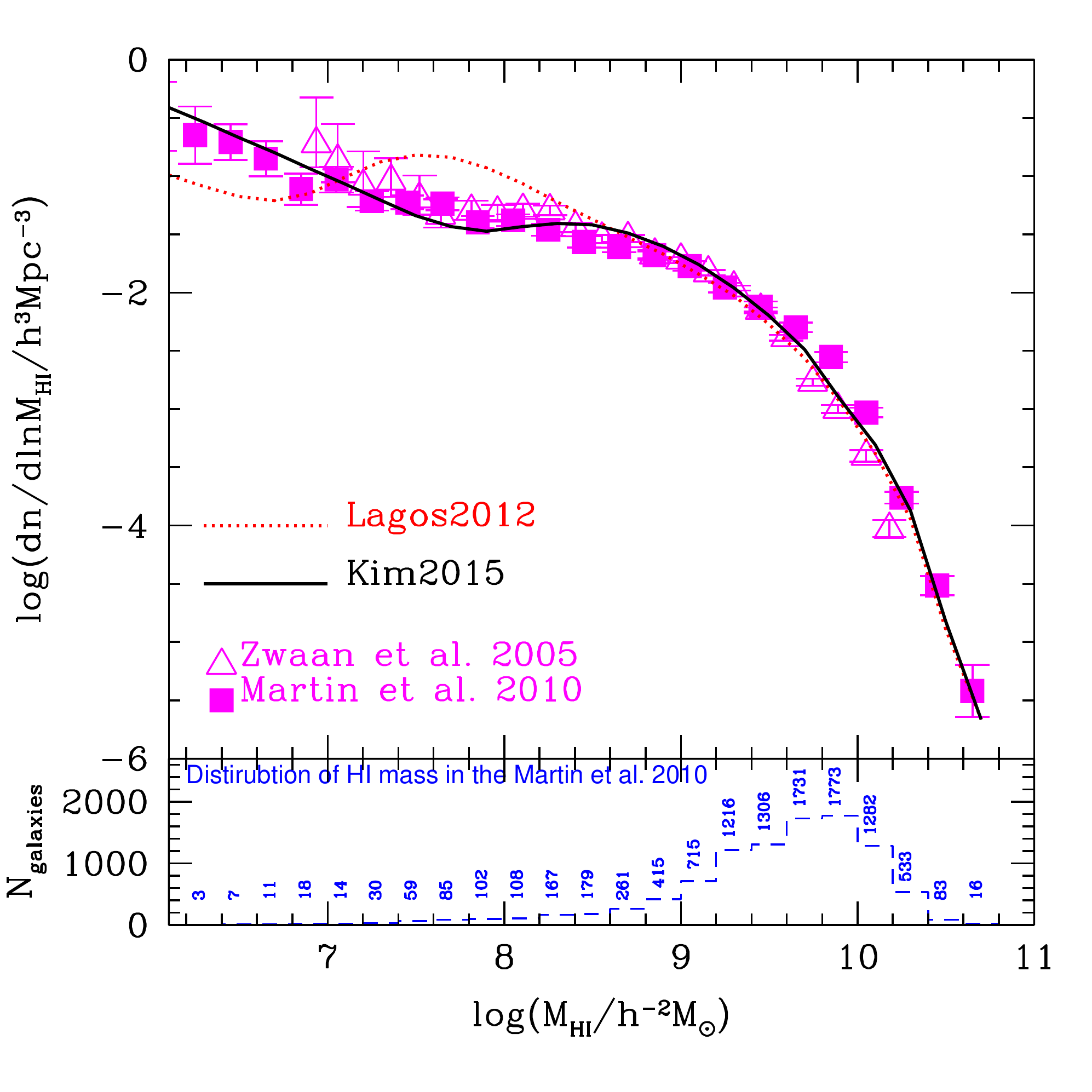}
    \includegraphics[width=8.cm]{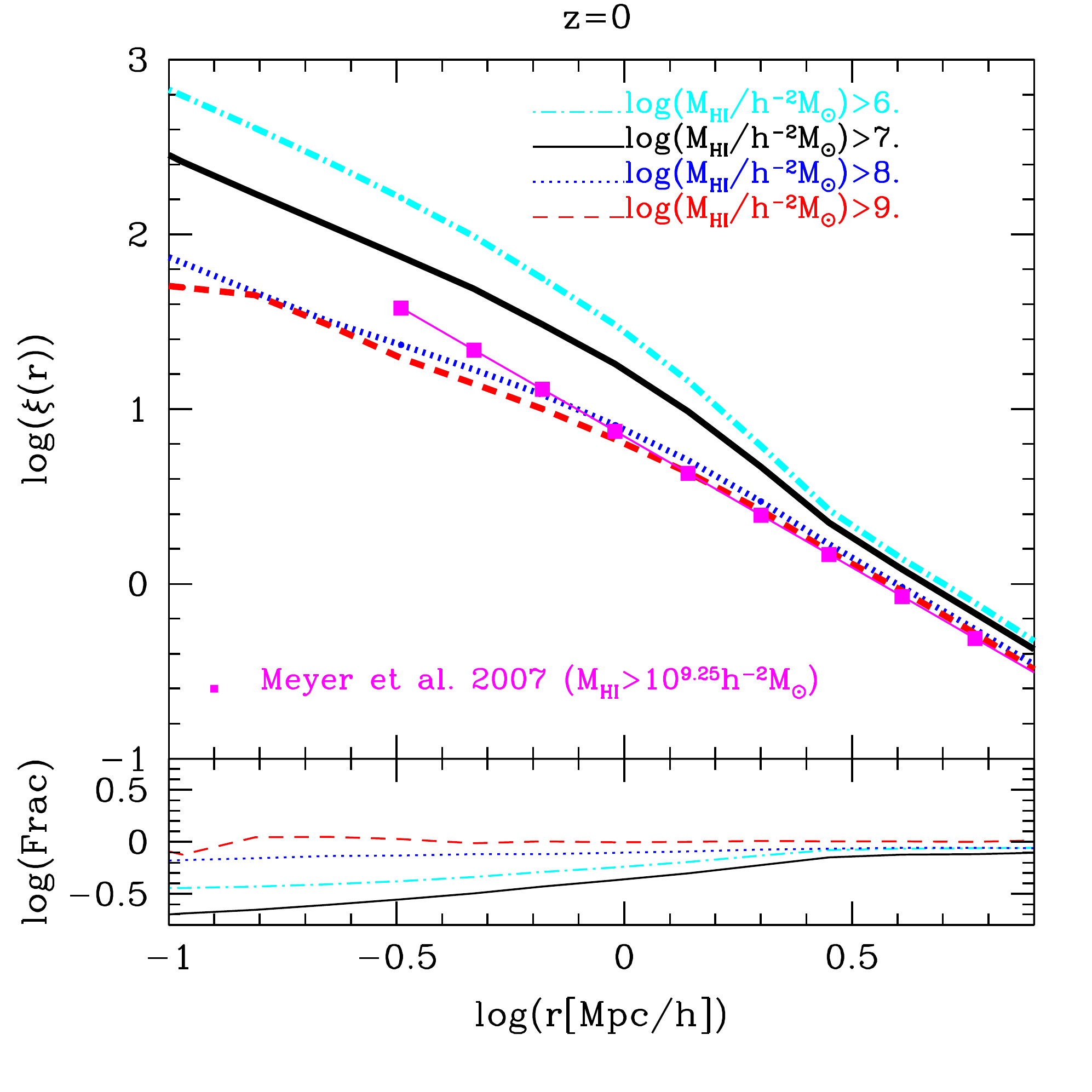}
  \caption{The left panel shows the HI mass function at $z$=0 predicted by the Lagos2012 (red dotted line) and the Kim2015 (black solid line) models {compared with local Universe observations from { HIPASS \citep[open triangles; cf.][]{zwaan.etal.2005}} and ALFALFA
    \citep[filled squares; cf.][]{martin.etal.2010}.} { The left bottom-sub panel shows the distribution of HI mass in the sample of \citet{martin.etal.2010} to estimate the HI mass function.} The right panel shows the correlation function of different HI mass threshold samples at $z$=0 for the Kim2015 model. The right bottom-sub panel shows the ratio between the predictions of the Lagos2012 model and the Kim2015 models for the same HI mass thresholds samples labelled in the right top-sub panel. The magenta filled squares connected by the solid line in the right top-sub panel show the measurement of \citet{Meyer2007} using their best fit $r_{0}$ and $\gamma$ values for galaxies with HI masses $>$10$^{9.25}h$$^{-2}$M$_{\odot}$ { from the local Universe.}
}
  \label{CFModelCOM}
\end{figure*}

The left panel of Fig.~\ref{CFModelCOM} shows the HI mass functions predicted by the Lagos2012 (red dotted line) and the Kim2015 (black solid line) models. The Kim2015 model applied a new scheme for photoionisation feedback in order to improve the agreement of the predicted low-mass end of the HI mass function with the estimation from the ALFALFA survey at z=0. { This work is based on the model in \cite{Kim2015HI} which improves the modelling of the lower mass end of HI mass function. Here, we also show the ALFALFA observations for comparison to the model, which extend to low masses than the HIPASS data.}

{ Surveys of HI-selected galaxies are not volume limited and low HI mass galaxies are visible over a smaller volume than high HI mass galaixes. The blue dashed line in the left bottom-sub panel shows the distribution of HI mass in the sample of \cite{martin.etal.2010} used to estimate the HI mass function.}

{ The sample used in \cite{martin.etal.2010} from the ALFALFA survey has complete source extraction for 40\% of its total sky area. \cite{martin.etal.2010} measure the HI mass function from a sample of $\sim$10,150 galaxies and apply both the 1/V$_{max}$ \citep{Schmidt1968} and the Two Dimensional Step-Wise Maximum Likelihood (2DSWML) methods \cite[]{zwaan.etal.2005}. In this work, we use the HI mass function estimated using the 1/V$_{max}$ method when comparing with our model predictions.\footnote{The 1/V$_{max}$ method treats each individual galaxy by weighting the galaxy counts by the maximum volume V$_{max,i}$. This strategy allows the inclusion of low-mass galaxies (only detected in the nearby Universe) in the same sample as rare high-mass galaxies (only found in larger volumes). In addition, the weights may be adjusted in order to correct for a variety of selection effects, large-scale structure effects, and missing volume within the dataset.} } 

The right panel of Fig.~\ref{CFModelCOM} shows the correlation functions for different HI mass thresholds in the Kim2015 model, compared with observations from \citet{Meyer2007} using the best fitting $r_{0}$ and $\gamma$ values for galaxies with HI masses $>$10$^{9.25}h^{-2}$M$_{\odot}$. The right bottom-sub panel shows the ratio between the predictions of the Lagos2012 and the Kim2015 models, for the same HI mass thresholds used in the second top-sub panel. These models differ in their treatment of photoionisation feedback as described in \S\ref{sec:model}. The right panel of {Fig.~\ref{CFModelCOM} shows that HI-selected galaxies down to} $M_{\rm HI}$$>$10$^{7}h^{-2}$M$_{\odot}$ could constrain the properties of photoionisation feedback \cite[which is also imprinted on the HI mass function,][]{Kim2015HI}, and the contribution of satellite galaxies to the distribution of HI-selected galaxies from HI clustering measurements through the $\gamma$ value of the correlation function.

Fig.~\ref{CFModelCOM} illustrates that observational HI clustering measurements which contain HI galaxies with $M_{\rm HI}$$<$10$^{8}h^{-2}$M$_{\odot}$ are necessary to constrain the modelling of galaxy formation, particularly how photoionisation affects small galaxies \cite[]{Kim2015HI}, and the contribution of satellite galaxies to the distribution of HI galaxies.

\section{Distribution of neutral hydrogen at high redshifts}
\label{sec:HIGHCLUSTERING}
We now investigate the predicted clustering of HI selected galaxies at high redshift in ongoing and future HI galaxy surveys.

\begin{figure*}
  \includegraphics[width=8.1cm]{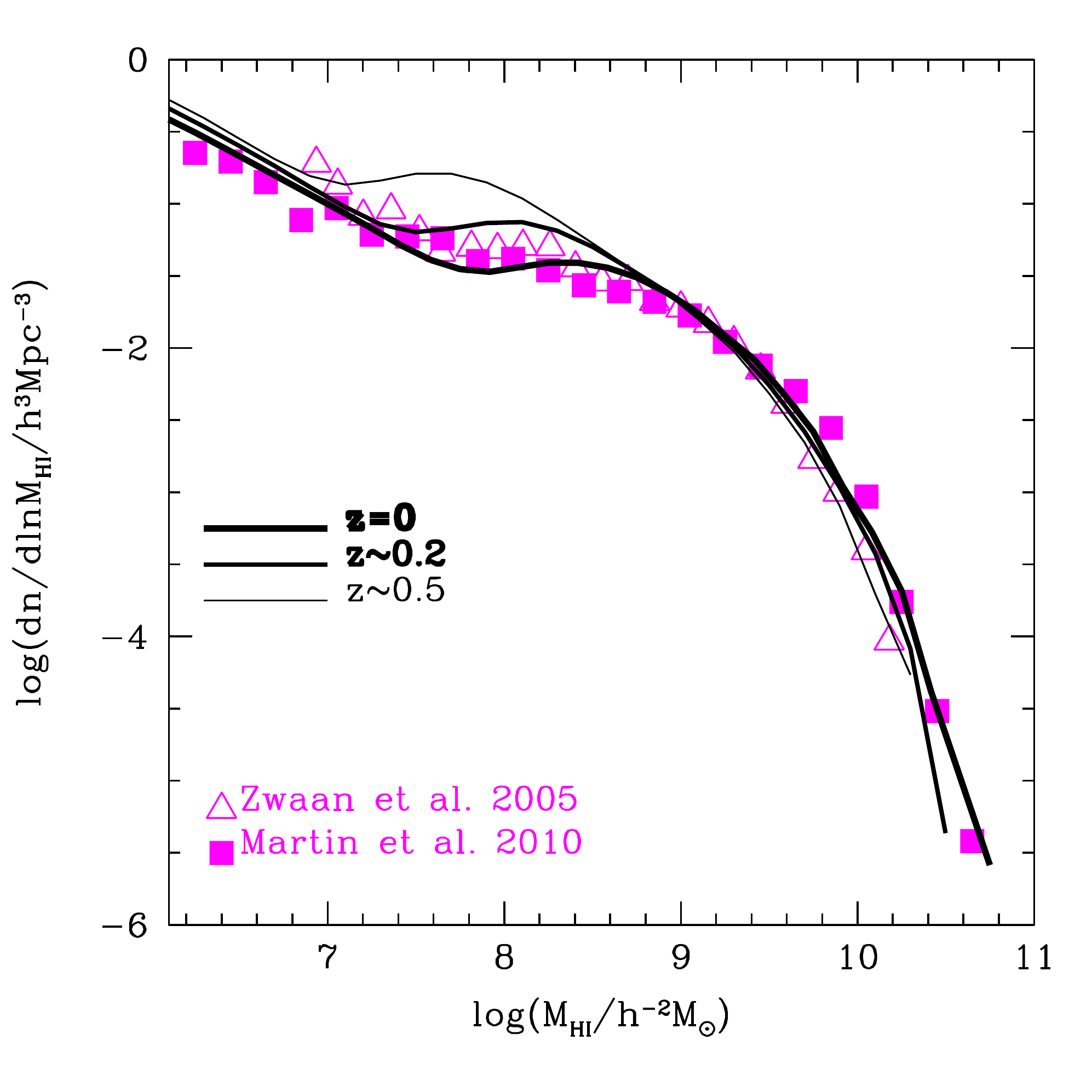}
    \includegraphics[width=8.cm]{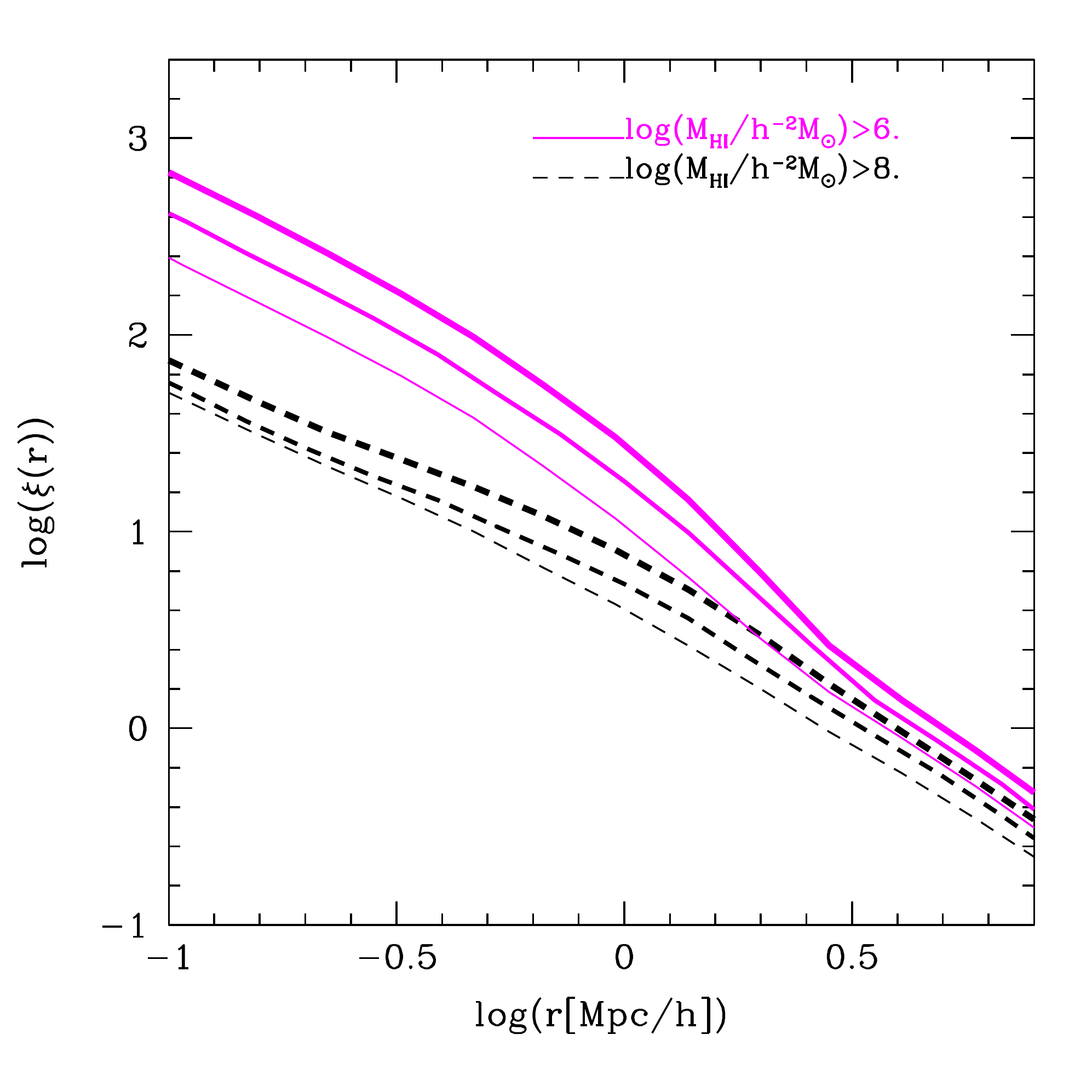}
  \caption{Left panel: the HI mass function predicted by the Kim2015 model at z=0, 0.2 and 0.5 as labelled, compared with local Universe observations from { HIPASS \citep[open triangles; cf.][]{zwaan.etal.2005}} and ALFALFA
    \citep[filled squares; cf.][]{martin.etal.2010}. Right panel: the real space correlation function predicted for two different HI mass thresholds samples at redshifts z=0, 0.2, and 0.5, as labelled in the left panel.}
 \label{HIMFZ0.5}
\end{figure*}

\cite{Power2010} showed that the HI mass function undergoes little evolution from $z$=1 to $z$=0 \cite[see also,][]{Kim2011, Lagos2011}. This is due to a competition between gas cooling, the transition from HI to H$_{2}$ and star formation, together with mass ejection by winds. This is a generic feature predicted by galaxy formation models for high HI mass galaxies \cite[see][]{Power2010}. However, this picture could break down for HI-poor galaxies, which reside in low mass { host} dark matter haloes. These haloes are more sensitive to photoionization feedback, which hampers gas cooling. Here, we investigate the evolution of the HI mass function. The left panel of Fig.~\ref{HIMFZ0.5} shows the predicted HI mass functions at $z$=0, 0.2, and 0.5 (corresponding to maximum redshifts expected to be observed by the WALLABY\footnote{http://askap.org/wallaby; the Widefield ASKAP L$-$band Legacy All$-$sky Blind surveY}(z$\sim$0.2), DINGO\footnote{http://internal.physics.uwa.edu.au/~mmeyer/dingo/welcome.html; Deep Investigations of Neutral Gas Origins}(z$\sim$0.5) and LADUMA\footnote{http://www.ast.uct.ac.za/laduma/node/6; Looking at the Distant Universe with the MeerKAT Array}(z$\sim$0.5)), together with the HI mass function from the ALFALFA survey at $z$=0 for reference. The largest difference between the HI mass functions at $z$=0.2 (or 0.5) and $z$=0 is predicted to be for galaxies with masses in the range 10$^{7}h^{-2}$M$_{\odot}<$M$_{\rm HI}<$10$^{8.5}h^{-2}$M$_{\odot}$. The circular velocities of the host dark matter haloes of these galaxies are between $\sim$ 30 km/s and $\sim$ 50 km/s in the Kim2015 model. These velocities correspond to galaxies which become affected by photoionization feedback between $z$=0.5 and $z$=0. 

The right panel of Fig.~\ref{HIMFZ0.5} shows the predicted correlation functions at $z$=0, 0.2 and 0.5 for two different HI mass thresholds (M$_{\rm HI}>$10$^{6}h^{-2}$M$_{\odot}$ and M$_{\rm HI}>$10$^{8}h^{-2}$M$_{\odot}$). The correlation functions at higher redshift have lower correlation length ($r_{0}$) and slope ($\gamma$) than at lower redshift for the same HI mass thresholds (see Table~\ref{Tabler0}). { The parameters $r_{0}$ and $\gamma$ were derived by minimising { $\chi^{2}$} of predicted correlation functions of the form shown in Eq.~\ref{powerlaw} over separations between 1 and 10$h^{-1}$Mpc and considering $\gamma$ values between 0 and 5. The separation range was chosen based on the approach in \cite{Meyer2007} and \cite{Martin2012} who used projected separations $<$10$h^{-1}$Mpc and \cite{papastergis2013} used projected separation between 0.5$h^{-1}$Mpc and 10$h^{-1}$Mpc.}

\begin{figure}
  \includegraphics[width=8.cm]{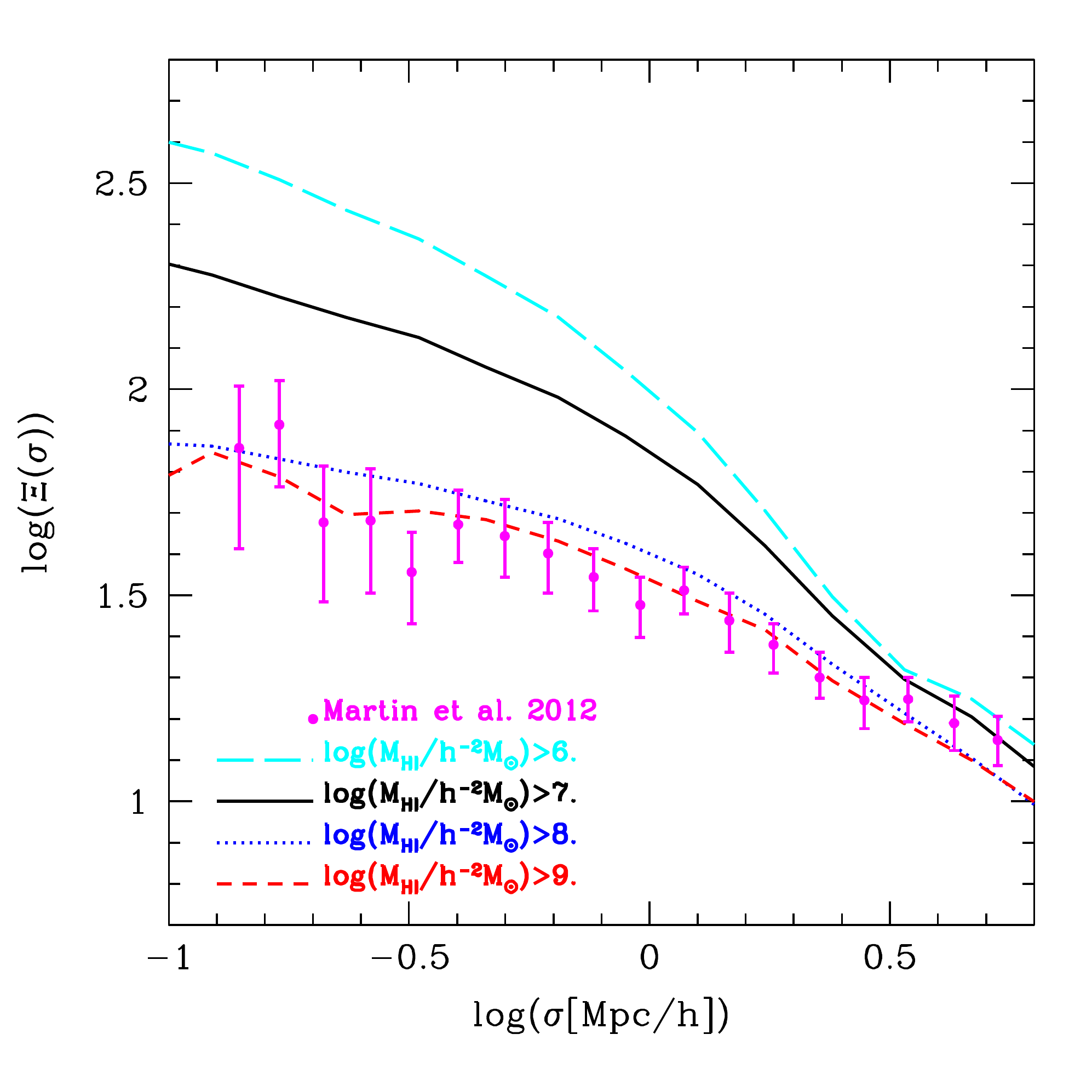}
   \caption{{ The projected galaxy correlation function at $z$=0 for different HI mass threshold samples from the model. $\sigma$ indicates the projected separations. The points with error bars show an observational estimate made from the ALFALFA
catalogue by \citet{martin.etal.2010} at local Universe.}}
  \label{2DCF}
\end{figure}

\begin{figure*}
  \includegraphics[width=8.cm]{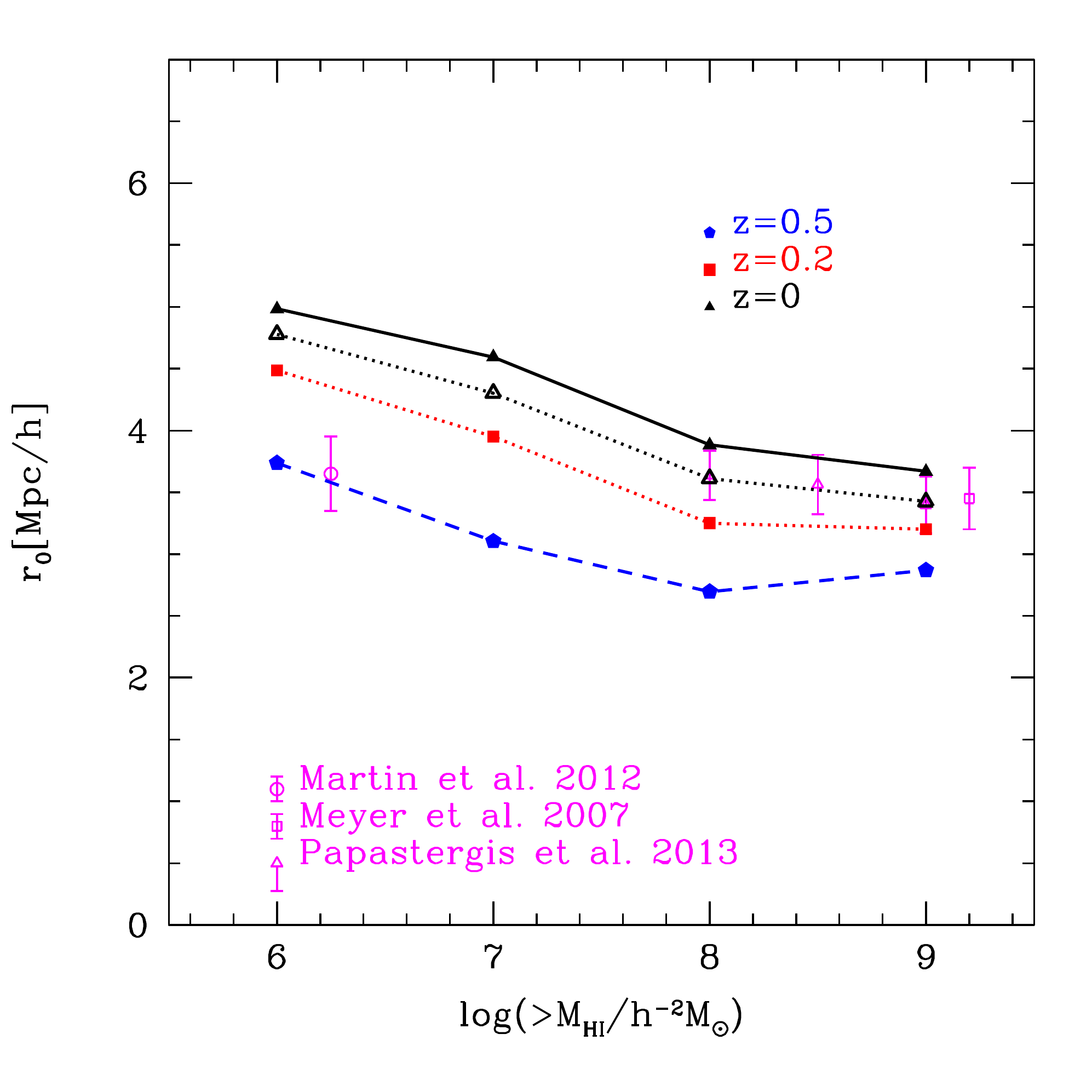}
  \includegraphics[width=8.cm]{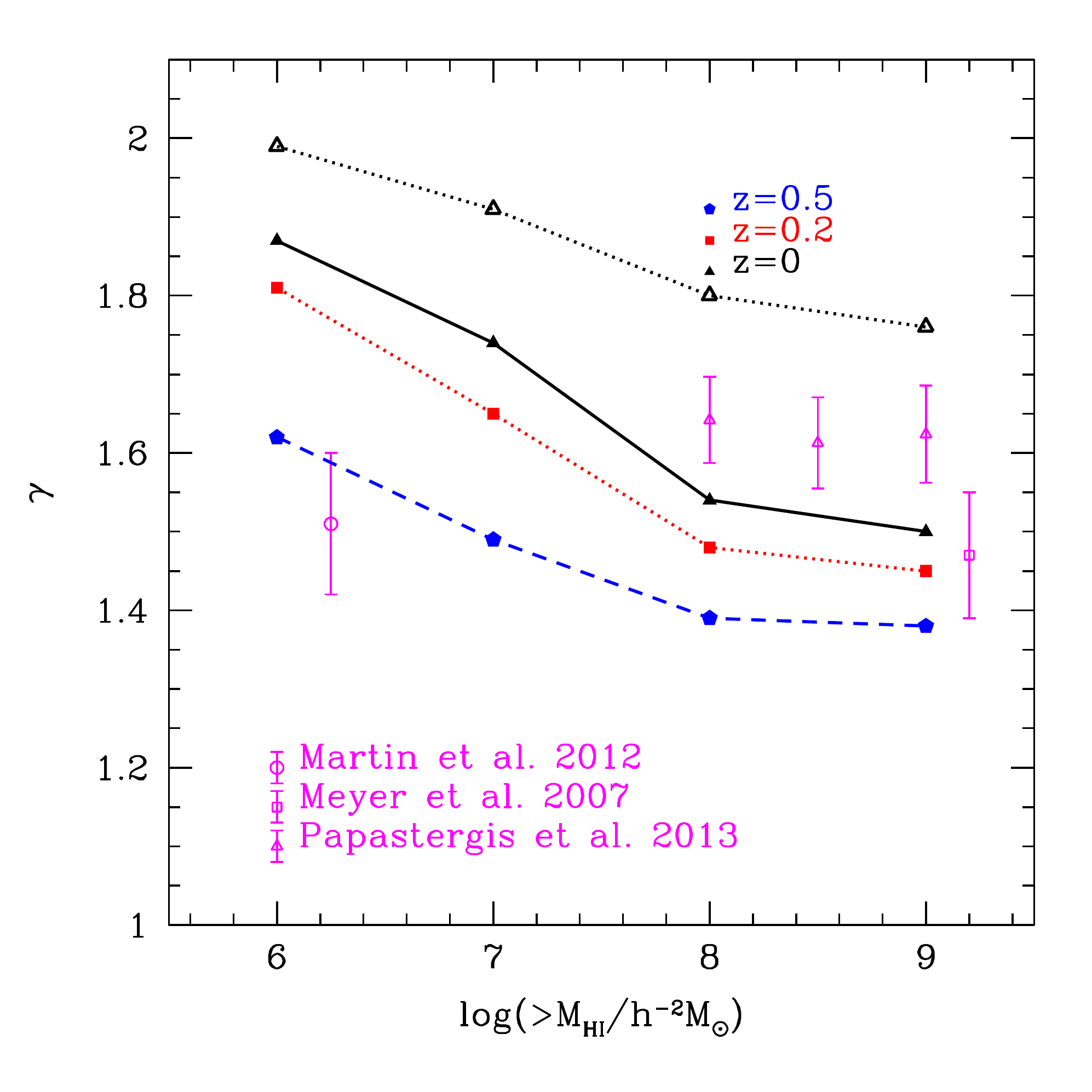}
  \caption{Left panel: The correlation length ($r_{0}$) of the correlation function for different HI mass thresholds at $z$=0, 0.2, and 0.5, as labelled. Right panel: The slope ($\gamma$) of the correlation function for different HI mass thresholds at $z$=0, 0.2, and 0.5, as labelled. For reference the symbols with errorbars (the standard deviation) show the best fit $r_{0}$ and $\gamma$ values from observations of HI-selected galaxies measured by \citet{Meyer2007}, \citet{Martin2012}, and \citet{papastergis2013} in the local Universe. { The open triangles show the r$_{0, {\rm power}}$ (left panel) and $\gamma_{\rm pow}$ (right panel) from the projected correlation function of the model following \citet{Meyer2007}}}
  \label{HIBP}
\end{figure*}

{ Figure~\ref{2DCF} shows the projected correlation functions in the model by projecting galaxy samples onto the 2D x-y plane within the z range $\sim$40$h^{-1}$Mpc by applying the distant-observer approximation. The points with error bars show an observational estimate made from the ALFALFA catalogue by \citet{martin.etal.2010} at local Universe. The projected correlation functions from the model do not show a power law shape. To see the effect of the power law assumption, we estimate the r$_{0}$ and $\gamma$ value following \cite{Meyer2007} using the projected correlation function from the model. We have also included the r$_{0}$ and $\gamma$ values from these results in table.~\ref{Tabler0} (in the square bracket) and Fig.~\ref{HIBP} for z=0 (connected open triangles by the dotted line) for comparison between the values from the real space correlation function and extracted values from the projected correlation function in the model.}

\begin{table*}
\caption{
The best fitting parameters (r$_{0}$[$h^{-1}$Mpc] and $\gamma$) in Eq.~\ref{powerlaw} for predicted correlation functions of different HI mass thresholds at $z$=0, 0,2 and 0.5 in the Kim2015 model. { In the square bracket at $z$=0, we show the r$_{0}$ and $\gamma$ value following \citet{Meyer2007} using the projected correlation function.}}
\label{Tabler0}
\begin{tabular}{ccccccc}
\hline
\hline
HI mass threshold & r$_{0}$[r$_{0,{\rm pow}}$] (z=0) &$\gamma$ [$\gamma_{\rm, pow}$] (z=0) &r$_{0}$ (z=0.2) &$\gamma$ (z=0.2 ) &r$_{0}$ (z=0.5) &$\gamma$ (z=0.5)\\
\hline
$>$10$^{6}h^{-2}$M$_{\odot}$ & 5.12 [4.91] &1.87 [1.99] &4.61&1.81&3.84&1.62\\
\hline
$>$10$^{7}h^{-2}$M$_{\odot}$ & 4.72 [4.42] &1.74 [1.91] &4.06&1.65&3.19&1.49\\
\hline
$>$10$^{8}h^{-2}$M$_{\odot}$  & 3.99 [3.71] & 1.54 [1.80]&3.34& 1.48& 2.77&1.39\\
\hline
$>$10$^{9}h^{-2}$M$_{\odot}$  & 3.77 [3.52] & 1.50 [1.76] & 3.29& 1.45& 2.95&1.38\\
\hline
\end{tabular}
\end{table*}

In the left panel of Fig.~\ref{HIBP}, we show variations on the predicted correlation length ($r_{0}$) at different redshifts and HI mass thresholds. In agreement with the $z$=0 predictions, we find that lower HI mass threshold samples show a larger correlation length than higher HI mass threshold samples at all redshifts in contrast to optically selected galaxy samples (c.f, red galaxy samples show an increasing $r_{0}$ at faint luminosities). This is because the contribution of galaxies hosted by high mass { host} dark matter haloes increases with decreasing HI mass threshold (see Fig~\ref{HankPHZ}).  

In the right panel of Fig.~\ref{HIBP}, we show the predicted correlation function slope ($\gamma$) for different redshifts and HI mass thresholds. $\gamma$ displays similar trends with redshift and HI mass threshold to $r_{0}$ (left panel of Fig.~\ref{HIBP}). Samples selected using lower HI mass thresholds show steeper correlation functions compared to samples using higher HI mass thresholds due to the contribution of satellite galaxies hosted by { host} dark matter haloes  of M$_{\rm halo}$$>$10$^{14}h^{-1}$M$_{\odot}$.  
In addition, the correlation length ($r_{0}$) and slope ($\gamma$) increase as redshift decreases from $z$=0.5 to $z$=0. This implies that galaxies at low redshift are more clustered than higher redshift galaxies selected above the same HI mass threshold, which is consistent with the hierarchical growth of structures. For comparison, the symbols with errorbars show the best fit value of $r_{0}$ (the left panel of Fig.~\ref{HIBP}) and $\gamma$ (the right panel of Fig.~\ref{HIBP}) estimated from observations in the local Universe of HI-selected for galaxies with $M_{\rm HI}$$>$10$^{9.25}h^{-2}$M$_{\odot}$ \citep{Meyer2007}, $M_{\rm HI}$$>$10$^{6.2}h^{-2}$M$_{\odot}$ \citep{Martin2012}, and $M_{\rm HI}$$>$ 10$^{8}$, 10$^{8.5}$, and 10$^{9}$$ h^{-2}$M$_{\odot}$ \citep{papastergis2013}. 

{ Note that the predictions are in reasonable agreement with the observations for M$_{\rm HI}>$10$^{8}h^{-2}$M$_{\odot}$. However, the model does
not reproduce the clustering estimated by \cite{Martin2012}
for samples with M$_{\rm HI}>$10$^{6.2}h^{-2}$M$_{\odot}$. The values 
of r$_{0}$ do not change significantly if the value of $\gamma$ is held 
fixed at the value found in the observational study. On the other hand, we
remind the reader that the observational clustering measurements
come from small volumes, which can be affected by cosmic variance. Although \cite{Martin2012} attempted to correct for the selection function of the sample, to account for
the fact that low HI mass galaxies are seen over a much smaller
volume than high HI mass galaxies, the cleanest approach for measuring galaxy clustering is to use volume limited samples. However, this is not yet feasible with
current HI surveys as this requires many galaxies to be omitted from the analysis \cite[see for example the volume limited samples
used in an optical galaxy survey by][]{norberg2001}. The assumption of a power law in the observed estimate of r$_{0}$ may also introduce a bias increasing the value of $\gamma$ and/or decreasing r$_{0}$ value.}
  
\section{Effect of galaxies with low HI masses on the signal for 21cm Intensity mapping}
\label{sec:INTENSITY}
As noted above, clustering studies of low HI mass-selected galaxies will be restricted to $z\sim0.5$, even within the deepest HI surveys planned (e.g., DINGO). 
An alternative way to study the clustering of galaxies down to low HI masses and at high redshifts (0.5$<$$z$$<$3) is through 21cm intensity mapping, which measures the fluctuation in the 21cm signal from unresolved HI-selected galaxies \citep{WL07}. This is a new strategy to economically map much larger volumes of the Universe, by measuring the collective 21cm emission from all galaxies within some volume set by the angular resolution of the telescope and the frequency interval samples. 

We predict the 21cm intensity mapping signal using the Kim2015 model. First, we divide the volume of the simulation into 256$^{3}$ cells (0.39Mpc/$h$ cell size corresponds to $\sim$ 1 arcmin at z=1, { but the value is arbitrary}). We then sum the mass of all HI in galaxies in each cell, and measure the resulting fluctuations of 21cm intensity.  

The power spectrum of 21cm brightness fluctuations is defined in \cite{Wyithe2010} as
\begin{equation}
P_{\rm 21cm}(k,z)\simeq T^{2}_{\rm b}x^{2}_{\rm HI}(z)\left< b(z) \right>^{2}P_{\rm DM}(k,z),
\end{equation}
where $T_{\rm b}$=23.8$\sqrt{(1+z)/10}$ mK. 
The term $<$$b(z)$$>$ is the HI mass-weighted halo bias, $P_{\rm DM}(k,z)$ is the power spectrum of the dark matter, and $x_{\rm HI}(z)$ is the fraction of neutral atomic hydrogen  $x_{\rm HI}(z)=\Omega_{\rm HI}(z)/0.76/\Omega_{\rm b}$ where $\Omega_{\rm HI}(z)$ is the neutral atomic hydrogen density and $\Omega_{\rm b}$ the baryon density of the Universe. { The predicted $\Omega_{\rm HI}(z)$ from the Kim2015 model agrees well with observations up to $z \sim$ 2 without any assumption about the relationship between HI mass and dark matter halo mass as was assumed in \cite{Bagla2010} \cite[]{Kim2015HI, Rhee2016}. The predicted HI mass increases monotonically with the halo mass above 10$^{12}$h$^{-1}$M$_{\odot}$ and can be well described by a power-law which is consistent with the zoom-in hydrodynamic simulation predictions in \cite{Villa2016}. \cite{Villa2016} only show this relation above M$_{halo}$$>$10$^{13}$M$_{\odot}$ because of the volume limit in their simulation.} Equivalently, we can predict the intensity mapping signal using the power spectrum of the HI mass of galaxies { which gives us information about the HI mass distribution from the model, with no assumptions about linear bias relative to the underlying dark matter distribution}, $P_{\rm HI}(k,z)$,
\begin{equation}\label{21PS}
P_{\rm 21cm}(k,z)=T^{2}_{\rm b}\Omega_{\rm HI}(z)/0.76/\Omega_{\rm b} \times P_{\rm HI}(k,z).
\end{equation}

\begin{figure}
  \includegraphics[width=8.cm]{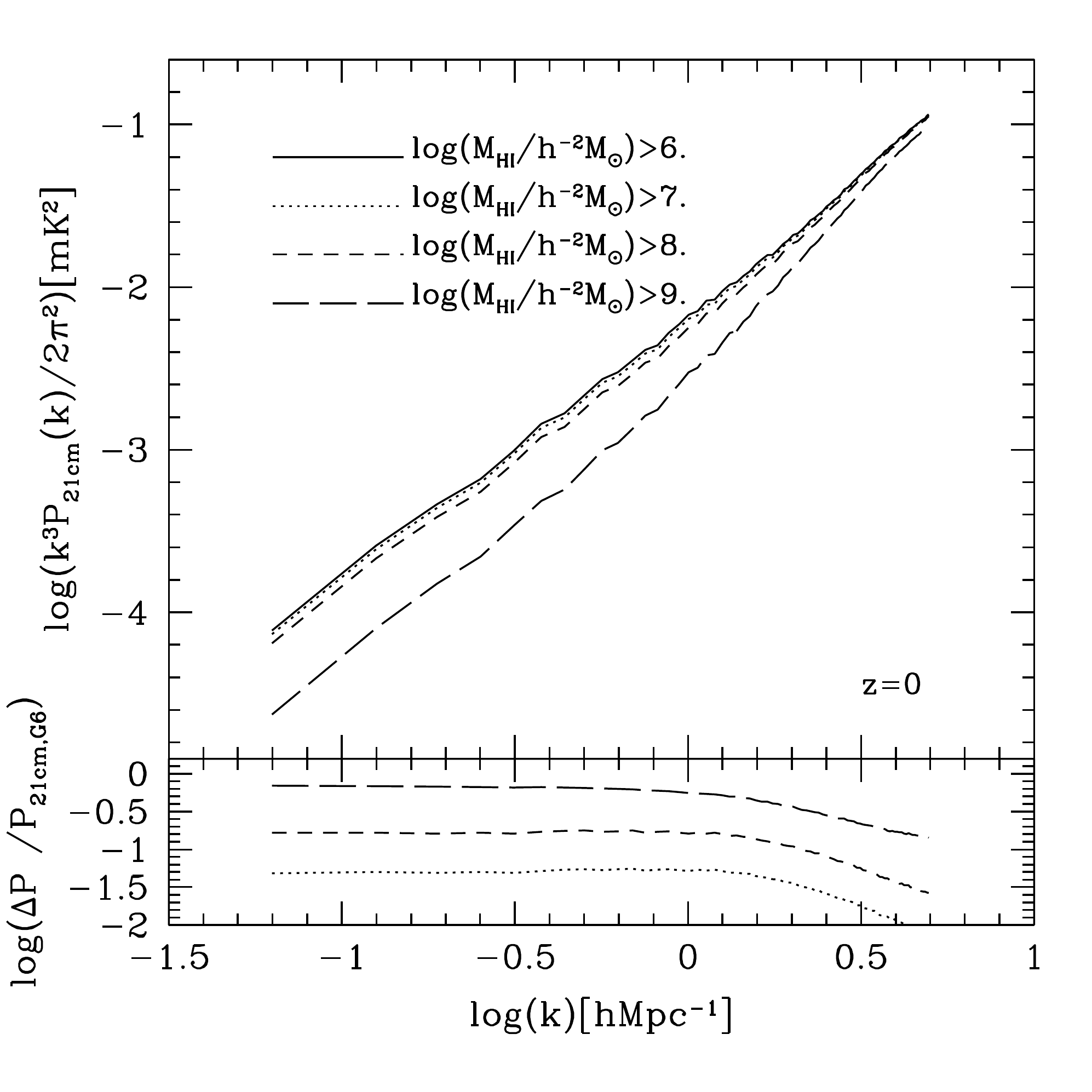}
  \caption{The predicted 21cm brightness temperature power spectrum as a function of HI mass threshold, as labelled, at $z$=0. We show the $\Delta P$ / $P_{\rm 21cm,G6}$ ratio in the bottom panel. $P_{\rm 21cm,G6}$ is the 21cm power spectrum using all galaxies with $M_{\rm HI}$$\ge$10$^{6}h^{-2}$M$_{\odot}$, and $\Delta P$ is the difference between power spectra of larger mass threshold and $P_{\rm 21cm,G6}$.}
  \label{HIPSZ0}
\end{figure}

Fig.~\ref{HIPSZ0} shows the predicted 21cm brightness temperature power spectrum predicted by the Kim2015 model for four different HI mass thresholds. The amplitude of the 21cm brightness temperature power spectrum decreases as the HI mass threshold increases, while the slope of the 21cm power spectrum increases with increasing HI mass threshold. We find that the 21cm intensity mapping power spectrum converges for M$_{\rm HI}$$>$10$^{7}h^{-2}$M$_{\odot}$. We also show the ratio $\Delta P$ / $P_{\rm 21cm,G6}$ in the sub-panel, where $\Delta P$ is the difference between a 21cm power spectrum using the galaxies above a given HI mass threshold and the 21cm power spectrum using all galaxies with $M_{\rm HI}$$\ge$10$^{6}h^{-2}$M$_{\odot}$ ($P_{\rm 21cm,G6}$). Note that we choose the $P_{\rm 21cm,G6}$ to compare with other HI mass thresholds because $M_{\rm HI}$$\sim$10$^{6}h^{-2}$M$_{\odot}$ is the minimum HI mass which can be trusted in the Kim2015 model based on the Millennium-II simulation (which has a minimum halo mass of few times 10$^{8}h^{-1}$M$_{\odot}$).

\begin{figure}
   \includegraphics[width=8.cm]{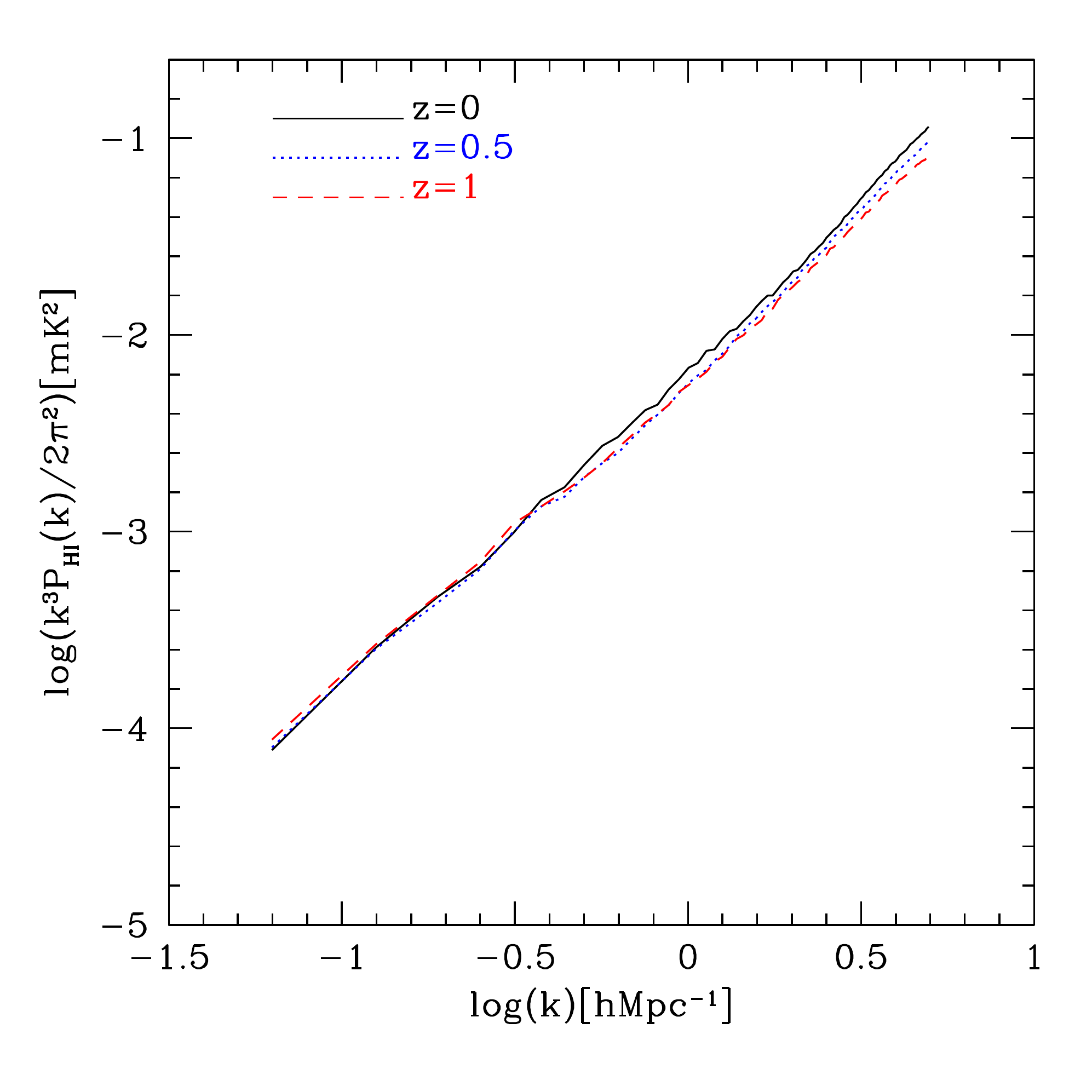}
  \caption{The predicted 21cm brightness temperature power spectrum { ($P_{\rm 21cm,G6}$)} for the Kim2015 model at z=0 (black solid line), 0.5 (blue dotted line), and 1 (red dashed line).}
  \label{HIPSZ1}
\end{figure}

We extend the prediction of the 21cm brightness temperature fluctuation to higher redshifts in Fig.~\ref{HIPSZ1}. There are { three} main factors driving the evolution of the 21cm brightness temperature power spectrum. { The first} is $\Omega_{\rm HI}$(z), which is the neutral hydrogen density. { The second} is the bias governed by the mass of { host} dark matter haloes hosting the HI-selected galaxies ($r_{0}$ and $\gamma$). { The third is HI mass contributions from each galaxy (i.e., HI masses) to HI mass intensity fluctuations.} The predicted value of $\Omega_{\rm HI}$(z) at $z$=1 is larger than at $z$=0.5 and $z$=0 \cite[see Fig~9 in][]{Kim2015HI}. However,  the clustering of { host} dark matter haloes hosting the HI-selected galaxies at $z$=0 is stronger, particularly at small separations (compare Fig.~\ref{CFModels} to  Fig.~\ref{HIMFZ0.5} and see Fig.~\ref{HIBP}). Thus, these { three main} competing drivers lead to different amplitudes and slopes for the predicted 21cm brightness temperature power spectra at different redshifts. { The third driver results in the largest difference between the HI galaxy correlation function results and 21cm intensity mapping power spectrum results.} 

\begin{figure}
\vspace{-0.4cm}
  \includegraphics[width=7.5cm]{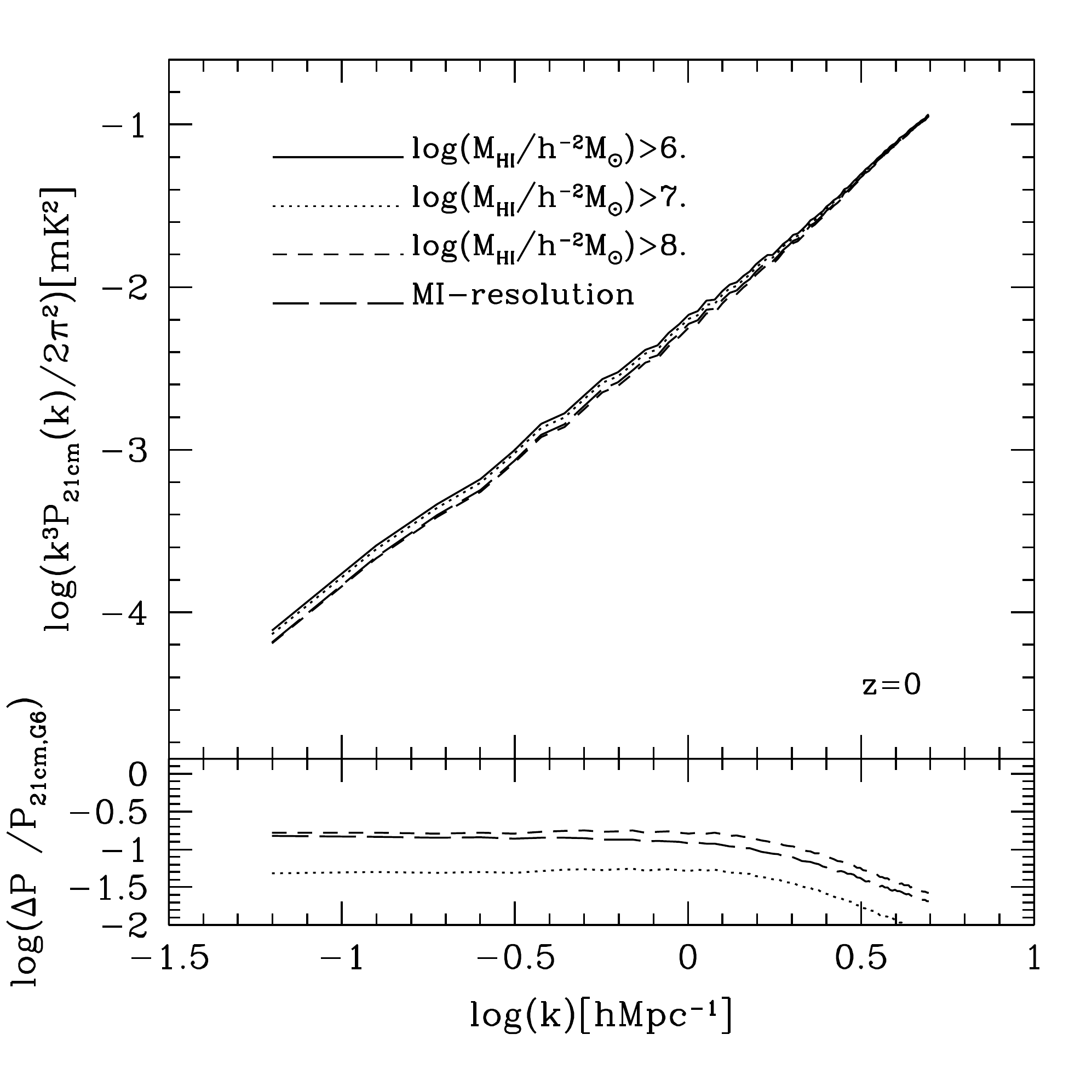}\vspace{-0.5cm}
  \includegraphics[width=7.5cm]{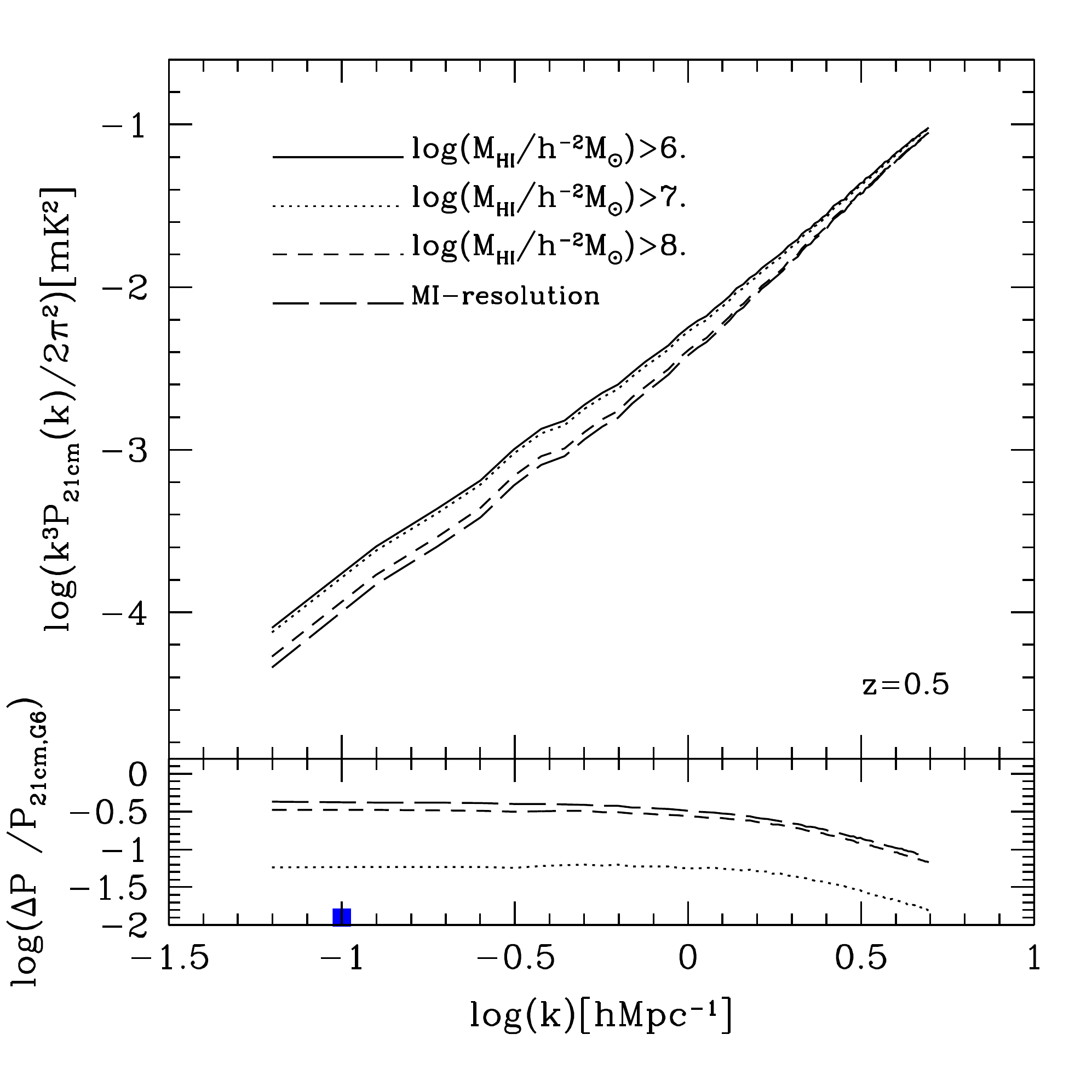}\vspace{-0.5cm}
  \includegraphics[width=7.5cm]{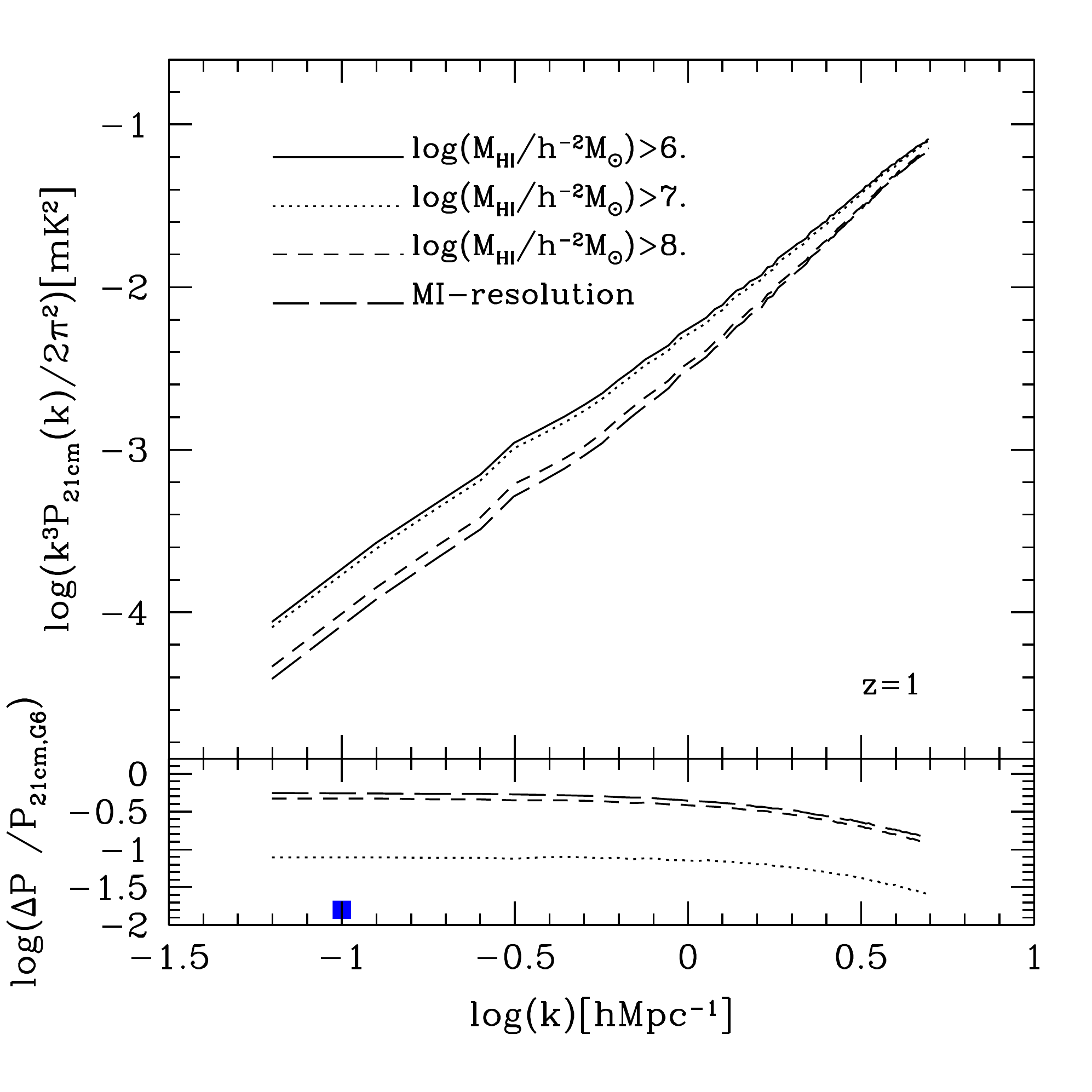}\vspace{-0.5cm}
  \caption{The predicted 21cm brightness temperature power spectrum for the Kim2015 model as a function of HI mass threshold at high redshifts, $z$=0 (top panel), 0.5 (middle panel), and z=1 (bottom panel). The long dashed line shows the measurement using the HI-selected galaxies in the host dark matter haloes which have the Millennium dark matter halo resolution imposed (labelled as MI-resolution). We show $\Delta P$ / $P_{21cm,G6}$, the fluctuation between thresholds and $P_{21cm,G6}$, in the sub-panels. The filled squares in the middle and bottom panels show the predicted constraint (noise over signal) at $k \sim 0.1$$h$Mpc$^{-1}$ expected for the SKA1-MID \citep[]{Santos2015}. }
  \label{HIPSZFRAC}
\end{figure}

We now focus on the contribution of unresolved galaxies. \cite{Power2010} showed that in the Millennium simulation, galaxies with M$_{\rm HI}$$<$10$^{8}h^{-2}$M$_{\odot}$ cannot be properly resolved (and so the completeness of the simulation drops below this HI mass). 
We explore the effect of unresolved galaxies on the 21cm signals from intensity mapping (Fig.~\ref{HIPSZFRAC}). We first ignore HI masses in galaxies that reside in the dark matter haloes { (both host and subhaloes)} less massive than 1.72$\times$ 10$^{10}h^{-1}$M$_{\odot}$ in the Millennium-II simulation (corresponding to the dark matter halo mass resolution of the larger Millennium simulation). The difference in 21cm brightness temperature fluctuations increases at higher redshifts (see long dashed lines in subpanels of Fig.~\ref{HIPSZFRAC}). This shows that low HI mass galaxies cannot be ignored if we aim to provide accurate predictions of the 21cm brightness temperature fluctuations at higher redshifts. 
  
\begin{table}
\caption{
The predicted dimensionless 21cm brightness temperature $\Delta$$^{2}$(k,z)[mk$^{2}$]=$k^{3}$$P_{\rm 21cm}$(k,z)/2/$\pi^{2}$ at $k_{p}$$\sim$0.126$h$Mpc$^{-1}$ of different HI mass thresholds and the Millennium simulation dark matter resolution case at $z$=0, 0,5 and 1 in the model.}
\label{Table21}
\begin{tabular}{lccc}
\hline
\hline
HI mass threshold & $\Delta$$^{2}$($k_{p}$,0) &$\Delta$$^{2}$($k_{p}$,0.5) &$\Delta$$^{2}$($k_{p}$,1) \\
(or Millennium resolution) & & &\\
\hline
$>$10$^{6}h^{-2}$M$_{\odot}$ & 2.375e-4 &2.5398e-4&2.674e-4\\
\hline
$>$10$^{7}h^{-2}$M$_{\odot}$ & 2.283e-4 &2.389e-4&2.463e-4\\
\hline
$>$10$^{8}h^{-2}$M$_{\odot}$  & 2.056e-4 & 1.698e-4&1.422e-4\\
\hline
  Millennium resolution & 2.049e-4& 1.479e-4& 1.197e-4\\
\hline
\end{tabular}
\end{table}

The GBT (Green Bank telescope) has pioneered the detection of the large scale structure using the technique of 21 cm intensity mapping \cite[]{Chang2010,Masui2013,Switzer2013}. In the near future, more sensitive experiments will improve upon this work, including BINGO, CHIME, and the SKA which will carry out 21cm intensity mapping measurements at different redshifts. We show the predicted sensitivity of SKA in the second and third panels of Fig.~\ref{HIPSZFRAC} at $k \sim 0.1h$Mpc$^{-1}$ based on calculations of \cite{Santos2015}. This predicted sensitivity of the SKA is one of main drivers of the effort behind the modelling of the HI content of galaxies down to low HI mass galaxies (M$_{\rm HI}$$<$10$^{7}h^{-2}$M$_{\odot}$) using a high enough resolution simulation { with a physically motivated galaxy formation model} to properly include these low HI mass galaxies.

\section{summary}
\label{sec:summary}

 We have predicted the clustering of HI galaxy samples as a function of HI mass threshold. The clustering of HI-selected galaxies is very similar for HI-selected galaxies with masses greater than 10$^{8}h^{-2}$M$_{\odot}$. This similarity can be explained in terms of the contribution of host dark matter haloes to the HI mass function. We estimate power-law fits for the predicted correlation functions and derive the correlation length $r_{0}$ and slope $\gamma$, as a function of redshift. We find that $r_{0}$ and $\gamma$ increase as the HI mass threshold decreases, which is the contrary to the expectations for optically selected galaxy samples. In addition, the predictions of clustering for different redshifts show that the correlation length and slope increase as redshift decreases from $z$=0.5 to $z$=0. 
We calculate the contribution of low HI mass galaxies to 21cm intensity mapping ($z$=0, $\sim$0.5, and $\sim$1). We find that it is important to model the HI mass function down to low HI masses of $\sim$ 10$^{7}h^{-2}$M$_{\odot}$ in order to correctly predict shape and amplitude of forthcoming observations of power spectra of 21cm intensity fluctuations. We also show that the importance of low HI mass galaxies to the 21cm brightness fluctuations increases at higher redshifts. Dark matter halo mass resolution in simulations must be sufficient (at least better than $\sim$10$^{10}{h}^{-1}$M$_{\odot}$) in order to properly predict the 21cm brightness temperature signal. 
With correct modelling of the distribution of HI-selected galaxies, 21cm intensity mapping surveys will provide a window to understanding of the small mass galaxy formation physics.

\section*{Acknowledgements}

{ We thank an anonymous referee for helpful comments on the manuscript.} H-SK is supported by a Discovery Early Career Researcher Awards from the Australian Research  Council (DE140100940). 
CP thanks Simon Driver and Aaron Robotham for helpful discussions. CP is supported by DP130100117, DP140100198, and FT130100041.
This work was supported by a STFC rolling grant at Durham. CL is funded by the ARC project DE150100618. 
CMB acknowledges receipt of a Research Fellowship from the  Leverhulme Trust. 
The calculations for this paper were performed on the ICC Cosmology Machine, 
which is part of the DiRAC Facility jointly funded by the STFC, the Large 
Facilities Capital Fund of BIS, and Durham University.  Part of the research 
presented in this paper was undertaken as part of the Survey Simulation 
Pipeline (SSimPL; {\texttt{http://www.astronomy.swin.edu.au/SSimPL/}). The 
Centre for All-Sky Astrophysics is an Australian Research Council Centre of 
Excellence, funded by grant CE110001020.

\newcommand{\noopsort}[1]{}

\bibliographystyle{mn2e}

\bibliography{HICFINAL}

\label{lastpage}
\end{document}